\documentclass[twocolumn,pre,superscriptaddress,floatfix,longbibliography]{revtex4-1}

\usepackage{graphics}
\usepackage{graphicx}
\usepackage{amsmath}
\usepackage{amssymb,xcolor}
\usepackage{subcaption} 

\definecolor{grisclair}{rgb}{0.6,0.6,0.6}

\newcommand{\beq}{\begin{equation}}
\newcommand{\ee}{\end{equation}}

\begin{document}

\title{Global linear stability of the bubble rising in the presence of a soluble surfactant}
\author{M. A. Herrada}
\address{E.T.S.I., Depto.\ de Ingenier\'{\i}a Aeroespacial y Mec\'anica de Fluidos, Universidad de Sevilla, Camino de los Descubrimientos s/n 41092, Spain}
\author{J. M. L\'opez-Herrera}
\address{E.T.S.I., Depto.\ de Ingenier\'{\i}a Aeroespacial y Mec\'anica de Fluidos, Universidad de Sevilla, Camino de los Descubrimientos s/n 41092, Spain}
\author{D. Fern\'andez-Mart\'{\i}nez}
\address{Depto.\ de Ingenier\'{\i}a Mec\'anica, Energ\'etica y de los Materiales and\\ 
Instituto de Computaci\'on Cient\'{\i}fica Avanzada (ICCAEx),\\
Universidad de Extremadura, E-06006 Badajoz, Spain}
\author{M. G. Cabezas}
\address{Depto.\ de Ingenier\'{\i}a Mec\'anica, Energ\'etica y de los Materiales and\\ 
Instituto de Computaci\'on Cient\'{\i}fica Avanzada (ICCAEx),\\
Universidad de Extremadura, E-06006 Badajoz, Spain}
\author{J. M. Montanero}
\address{Depto.\ de Ingenier\'{\i}a Mec\'anica, Energ\'etica y de los Materiales and\\ 
Instituto de Computaci\'on Cient\'{\i}fica Avanzada (ICCAEx),\\
Universidad de Extremadura, E-06006 Badajoz, Spain}

\begin{abstract}
We study the stability of the bubble rising in the presence of a soluble surfactant numerically and experimentally. For the surfactant concentration considered, the Marangoni stress almost immobilizes the interface. However, the non-zero surface velocity is crucial to understanding the surfactant behavior. The global linear stability analysis predicts the transition to an oblique path above the threshold of the Galilei number (the bubble radius). This transition is followed by the coexistence of stationary and oscillatory instabilities as the Galieli number increases. These predictions agree with the experimental observations without any fitting parameters. The bubble deformation, hydrostatic pressure variation, and perturbed viscous stress are evaluated. The velocity field perturbation causes a destabilizing vortex in the rear of the bubble. The perturbed viscous stress produces a torque opposing this vortex. The torque significantly decreases above the critical Galilei number, which may constitute the origin of instability. The linear stability analysis and the experiments were conducted for Surfynol, which can be regarded as a fast surfactant. Our experiments show the considerable differences between the rising of bubbles in the presence of this surfactant and a regular one. 
\end{abstract}

\maketitle

\section{Introduction}

The rising of a bubble in still water is a paradigmatic problem that has been analyzed for several centuries and continues to capture the attention of many researchers today \citep{S56,SSSW08,TSG15,CTMM16,CMMT16,KJLS21,BFM23}. The most exciting phenomenon when a bubble rises in water is the path instability occurring for radii exceeding a critical value, which gives rise to a rich and complex phenomenology. 

Global linear stability is probably the best approach to elucidate the mechanisms responsible for this phenomenology \citep{B19b,HE23,BFM23}. The first studies considered the stability of the flow past fixed solid bodies with spheroidal \citep{NA93,YP07,ERFM12,TMF13} and more realistic fore-and-aft axisymmetric \citep{CBM13} shapes. The importance of interplay between the flow and an inertialess mobile solid body was pointed out in subsequent works \citep{TMF14,CTMM16,BSFM24}. Some studies have also considered the bubble deformation to predict accurately the critical radius and the path oscillation frequency at the marginal stability \citep{ZD17,B19b,HE23,BFM23}. The instability corresponds to a supercritical Hopf (oscillatory) bifurcation with an azimuthal number $m=1$. The most accepted explanation of the path instability is based on the interplay between the wake and the bubble motion as a solid rigid \citep{B19b,BFM23,BSFM24}. \citet{HE23} have claimed that bubble deformation plays a significant role. Which oscillatory path is observed experimentally depends on the initial conditions or is selected by nonlinear effects. Direct numerical simulations \citep{CBM13,TSG15,CMMT16} and experiments \citep{S56,D95} have determined that this bifurcation typically leads to a zigzag motion.

The rise of a bubble in perfectly clean water is practically an idealization of the real problem, in which surfactant is always present because it is added either on purpose or unintentionally as impurities. When a bubble is released in a liquid bath containing surfactant, the surfactant molecules adsorb onto the free surface during the bubble rising substantially changing the bubble dynamics \citep{YI87,RKF96,FD96,ZF01,TKMB04,T05,AOV05,KJ05,TM11,LWZGZXLL22,PJF23}, even at tiny surfactant concentrations \citep{RVCMLH24,FCLHM25}. The bubble rising in the presence of a surfactant is a complex phenomenon in which fluid-dynamic and physicochemical processes are coupled. Interface expansion/compression, sorption kinetics, and convection over the bubble surface compete to establish the so-called dynamic adsorption layer \citep{DML98,DKGLKMM15,UGGLGM16,ZMRABSF23}. Although the explanation of the surfactant effect on the bubble shape and velocity is well accepted, only a few quantitative comparisons between numerical simulations and experiments have been conducted \citep{LM01,PPM06,TUWM03,T05,RVCMLH24,FCLHM25}. 

Surfactants dissolved in water are known to destabilize the path of a rising bubble. Helical and zigzag trajectories have been observed in direct numerical simulations when the bubble radius exceeds a critical value \citep{PWMB18}. Most experiments have been conducted for concentrations well above the critical one for a given bubble radius, which prevents one from analyzing the instability transition. \citet{TTM14} systematically analyzed how the surfactant enhances the path instability. They observed the transition from a helical motion to a zigzag trajectory, something not reported previously in the case of purified water. To the best of our knowledge, the global linear stability of a bubble rising in the presence of surfactant has not yet been conducted despite its relevance at the fundamental and practical levels. The goal is to determine the critical bubble radius for a given surfactant concentration and elucidate the mechanism responsible for the instability. 

Consider the time evolution of the surface coverage when an initially clean spherical bubble remains at rest in a liquid bath containing a surfactant at a very low concentration. In this case, the (linear) Henry model $\Gamma=k_a c/k_d$ relates the surface $\Gamma$ and volumetric $c$ surfactant concentrations at equilibrium \citep{MS20}. Here, $k_a$ and $k_d$ are the adsorption and desorption constants, respectively. In the sorption-limited case, the characteristic time scale for the surfactant adsorption is $\tau_k=1/k_d$. The corresponding scales in the diffusion-controlled limit are $\tau_D=L_d^2/{\cal D}$ for $L_d\ll R$ and $\tau_D=L_d R/{\cal D}$ for $\L_d\gg R$ ($L_d=k_a/k_d$ is the depletion length and $R$ is the bubble radius). The Damkohler number $\text{Da}=\tau_D/\tau_k$ indicates whether surfactant adsorption is controlled by diffusion ($\text{Da}\gg 1$), sorption kinetics ($\text{Da}\ll 1$), or both ($\text{Da}\sim 1$). This number typically takes values much greater than unity (for instance, $\text{Da}\sim 10$ and $10^2$ for a bubble 1 mm in radius loaded with SDS and Triton X-100, respectively). Therefore, surfactant adsorption in a bubble at rest is typically limited by diffusion. In this case, the sorption constants enter the problem only through the depletion length $L_d$ \citep{MS20}. 

\begin{figure}
\vspace{0.cm}
\begin{center}
\resizebox{0.43\textwidth}{!}{\includegraphics{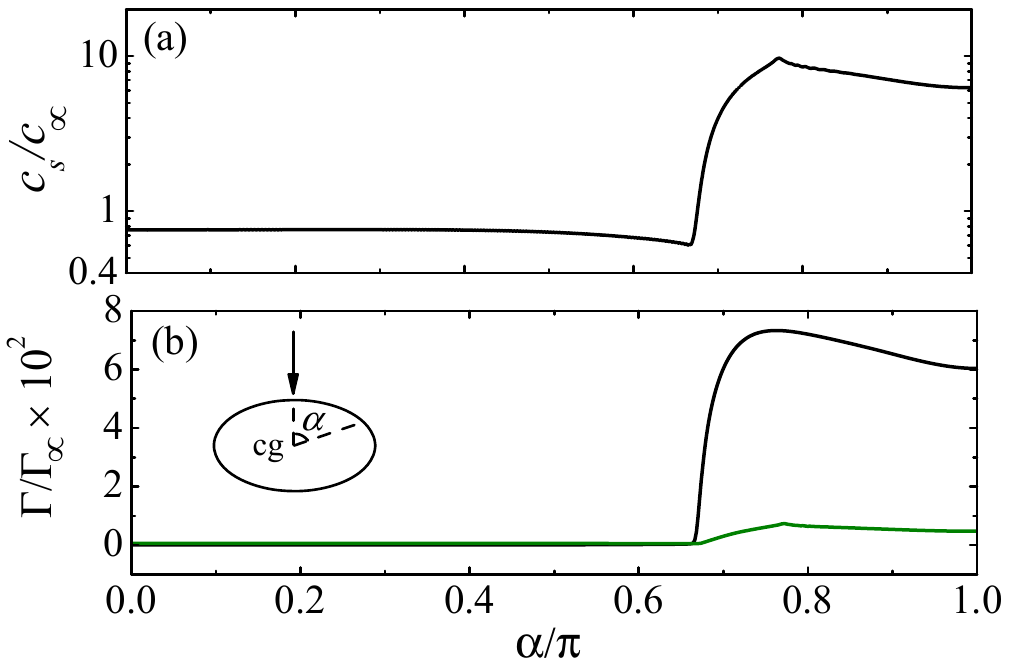}}
\end{center}
\caption{Simulation results calculated by \citet{RVCMLH24} for a bubble 0.66 mm in radius rising in SDS aqueous solution at the concentration $10^{-4}$ times the critical micelle concentration. (a) Surfactant volumetric concentration at the bubble surface, $c_s$, in terms of the bath concentration, $c_{\infty}$. (b) Surfactant surface concentration, $\Gamma$, in terms of the maximum packing concentration, $\Gamma_{\infty}$. The black lines are the simulation results, while the green line in panel (b) is the value obtained from the equilibrium equation $\Gamma=L_d c_s$.}
\label{fast}
\end{figure}

In a rising bubble, convection collaborates with diffusion to transport the surfactant molecules across the liquid bath. This implies that sorption kinetics becomes a limiting mechanism even for large values of Da. For most surfactants, the bubble trajectory depends on the sorption constants separately (not only through $L_d$)  \citep{FCLHM25}, and the equilibrium approximation $\Gamma=k_a c_s/k_d$ ($c_s$ is the concentration at the interface) cannot be considered (Fig.\ \ref{fast}). Only fast surfactants, such as Surfynol \citep{VSQCC24}, are expected to overcome convection so that sorption kinetics can be regarded as instantaneous, even for a rising bubble. In that case, the equilibrium relationship $\Gamma(c_s)$ approximately verifies, and the sorption constants enter the problem only through their ratio $L_d$, which can be determined from surface tension measurements at equilibrium. 

When a non-fast surfactant, such as SDS, dissolves at a low concentration, it accumulates at the bubble surface, forming an extremely thin diffusive surface boundary layer, which produces a Marangoni stress three orders of magnitude larger than the tangential viscous stress in a surfactant-free bubble \citep{RVCMLH24}. The diffusive boundary layer results from the surfactant accumulation in the bubble rear. Convection ``smashes the surfactant against the bubble south pole" (the surfactant molecules can leave the interface there only through desorption). This phenomenon is less likely to occur with a fast surfactant. In this case, the surface concentration is at equilibrium with the volumetric one at that point. Therefore, large surface concentration gradients entail large volumetric concentration gradients in the streamwise direction. However, the surfactant molecules are convected across the bulk without running into a ``topological obstacle" (the bubble south pole in the case of the surfactant molecules adsorbed onto the interface), which hinders the growth of large streamwise gradients. We conclude that thin diffusive surface boundary layers are less likely to occur with a fast surfactant. In other words, one expects stagnation caps to appear under more stringent parameter conditions with fast surfactants. In most experimental realizations, a fast surfactant is expected to be smoothly distributed over the interface, facilitating the global stability analysis. 

We study the path stability of a bubble rising in water with surfactant both experimentally and numerically. The experimental analysis describes the peculiarities of bubble rising in the presence of a fast surfactant by comparing our results with those of SDS. The numerical study presents the first global stability analysis of a bubble rising in the presence of a surfactant. This analysis allows us to predict the bubble behavior observed in our experiments. We examine the perturbations responsible for the primary instability. The similarities and differences between the surfactant-covered bubble and a bubble with a rigid surface are elucidated.   

\section{Methods}
\label{sec2}

\subsection{Governing equations}
\label{sec21}


Consider a bubble of radius $R=[3V/(4\pi)]^{1/3}$ ($V$ is the bubble volume) rising in a liquid of density $\rho$ and viscosity $\mu$. The surface tension of the clean interface is $\gamma_c$, while the gravitational acceleration is $g$. We dissolve a surfactant in the liquid at the concentration $c_{\infty}$. In the framework of the model considered here, the relevant surfactant properties are the volumetric diffusion coefficient ${\cal D}$, the surface diffusion coefficient ${\cal D}_S$, the depletion length $L_d$, and the maximum packing density $\Gamma_{\infty}$.

\begin{figure}
\vspace{0.cm}
\begin{center}
\resizebox{0.35\textwidth}{!}{\includegraphics{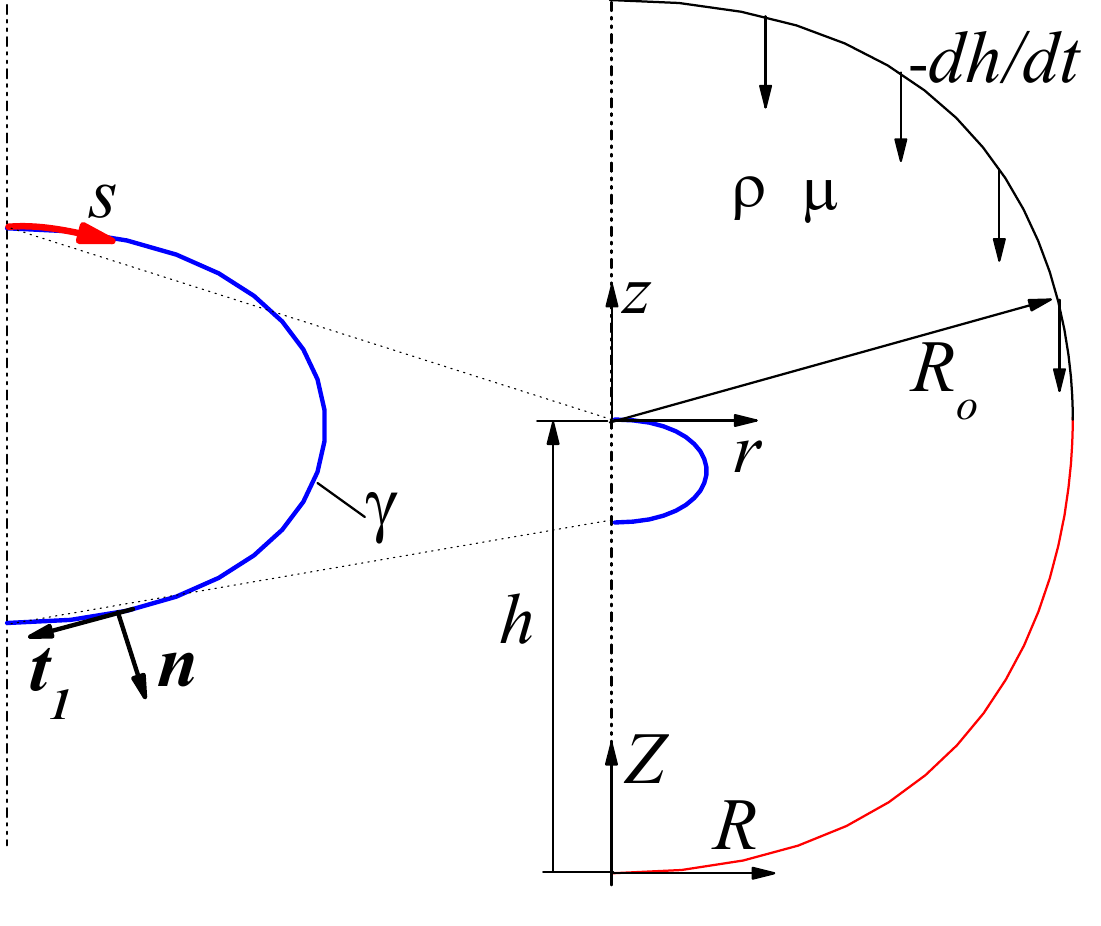}}
\end{center}
\caption{Sketch of the numerical domain. The blue and red outer boundaries correspond to the inlet and non-reflecting boundary conditions, respectively.}
\label{sketch}
\end{figure}

It is well known that the gas dynamics inside the bubble have negligible effects. The hydrodynamic equations are solved in the liquid phase using a cylindrical system of coordinates $(r,\theta,z)$ whose origin solidly moves with the bubble's upper point (Fig.\ \ref{sketch}). The continuity and momentum equations are
\begin{equation}
\label{equ1}
\boldsymbol{\nabla}\cdot\mathbf{v}=0, 
\end{equation}
\begin{equation}
\rho\frac{D\mathbf{v}}{Dt}=-\rho \left(g+\frac{d^2h}{dt^2}\right)\, \mathbf{e_z}+\boldsymbol{\nabla}\cdot\boldsymbol{\sigma},
\end{equation}
where $\mathbf{v}=v_{r}\mathbf{e_r}+v_{\theta}\mathbf{e_{\theta}}+v_{z}\mathbf{e_z}$ is the velocity field, $\mathbf{e_r}$, $\mathbf{e_{\theta}}$, and $\mathbf{e_z}$ are the unit vectors along $r$, $\theta$, and $z$, respectively, $D/Dt$ is the material derivative, $h=Z-z$ is the vertical position of the bubble's upper point (Fig.\ \ref{sketch}),
\begin{equation}
\boldsymbol{\sigma}=-p {\bf I}+\boldsymbol{\tau}
\end{equation}
is the stress tensor, $p$ is the hydrostatic pressure, ${\bf I}$ is the identity matrix, and
\begin{equation}
\boldsymbol{\tau}=\mu\left[\boldsymbol{\nabla}{\bf v}+\left(\boldsymbol{\nabla}{\bf v}\right)^T\right]
\end{equation}
is the viscous stress tensor. 

The interface is parametrized in terms of the meridional arc length $s$ as $r_i=F(s,\theta,t)$ and $z_i=G(s,\theta,t)$, where $(r_i,z_i)$ is the interface location. Here, $s$ is the arc length divided by the value corresponding to the bubble's rear point ($0\leq s\leq 1$) (Fig.\ \ref{sketch}). The kinematic compatibility condition reads
\begin{eqnarray}
&&\left(v_r-\frac{\partial F}{\partial t}\right)\frac{\partial G}{\partial s}-\left(v_z-\frac{\partial G}{\partial t}\right)\frac{\partial F}{\partial s}\nonumber\\
&&+\frac{v_{\theta}}{F}\left(\frac{\partial F}{\partial s}\frac{\partial G}{\partial \theta}-\frac{\partial F}{\partial \theta}\frac{\partial G}{\partial s}\right)=0.\label{kinematic}
\end{eqnarray}

The equilibrium of normal and tangential stresses on the interface leads to the equations
\begin{equation}
\label{bcb}
\mathbf{n}\cdot{\boldsymbol \sigma}\cdot \mathbf{n}=\gamma \kappa,\quad
\mathbf{t_1}\cdot{\boldsymbol \sigma}\cdot \mathbf{n}=\tau_{\text{Ma}}^{(1)}, \quad 
\mathbf{t_2}\cdot{\boldsymbol \sigma}\cdot \mathbf{n}=\tau_{\text{Ma}}^{(2)}
\end{equation}
where 
\begin{eqnarray}
{\bf n} &=& \frac{G_s {\bf e}_r-F_s{\bf e}_z+(F_s G_\theta-F_\theta G_\theta)/F\, {\bf e}_\theta }{[G_s^2+F_s^2+((F_s G_\theta-F_\theta G_\theta)/F)^2]^{1/2}},\nonumber \\
{\bf t}_1 &=& \frac{G_s{\bf e}_z+ F_s{\bf e}_r }{(F_s^2+G_s^2)^{1/2}},\nonumber\\
{\bf t}_2&=& {\bf n}\times {\bf t}_1,
\end{eqnarray}
are the normal and tangential unit vectors to the interface (the subscripts $s$ and $\theta$ denote the partial derivatives with respect to $s$ and $\theta$, respectively), $\gamma$ is the local value of the interfacial tension, $\kappa={\boldsymbol \nabla}_S\cdot \mathbf{n}$ is (twice) the mean curvature, and ${\boldsymbol\nabla_S}$ is the tangential intrinsic gradient along the free surface. In addition, 
\begin{equation}
\tau_{\text{Ma}}^{(1)}=-{\boldsymbol \nabla}_S\gamma\cdot \mathbf{t_1}, \quad  \tau_{\text{Ma}}^{(2)}=-{\boldsymbol \nabla}_S\gamma\cdot \mathbf{t_2} 
\end{equation}
are the tangential components of the Marangoni stress. We neglect the viscous surface stresses in Eqs.\ (\ref{bcb}) because the shear and dilatational viscosities of the surfactant monolayer are very small \citep{PRHEM20}.

We restrict ourselves to low surfactant concentrations, which implies that the surfactant is present as monomers. The monomer volumetric concentration $c({\bf r},t)$ in the outer phase is calculated from the conservation equation \citep{CMP09,KB19}
\begin{equation}
\label{c11}
\frac{\partial c}{\partial t}+\mathbf{v}\cdot \boldsymbol{\nabla}c={\cal D}\boldsymbol{\nabla}^2 c.
\end{equation}

Now, we assume that the sorption kinetics is much faster than any hydrodynamic process. Therefore, the sublayer and the interface are at equilibrium at any interface point and instant, which, according to the Langmuir adsorption isotherm \citep{MS20}, implies that
\begin{equation}
\label{km}
\frac{\Gamma/\Gamma_{\infty}}{1-\Gamma/\Gamma_{\infty}}=\frac{L_d\,  c_s}{\Gamma_{\infty}}.
\end{equation}
Here, $c_s$ is the bulk surfactant concentration evaluated at the interface and $\Gamma$ is the surfactant surface concentration (the surface coverage, measured in mols per unit area). 

The surfactant surface concentration $\Gamma$ verifies the advection-diffusion equation \citep{CMP09}
\begin{equation}
\label{ad}
\frac{\partial \Gamma}{\partial t}+{\boldsymbol \nabla}_S\cdot(\Gamma {\bf v}_S)+\Gamma({\boldsymbol \nabla}_S\cdot {\bf n})({\bf v}\cdot\mathbf{n})={\cal D}_S\nabla_S^2 \Gamma+{\cal D}\left. \nabla c\right|_n,
\end{equation}
where ${\bf v}_S={\boldsymbol {\sf I}}_S{\bf v}$ is the (two-dimensional) surface velocity, ${\boldsymbol {\sf I}}_S={\boldsymbol {\sf I}}-\mathbf{n}\mathbf{n}$ is the tensor that projects any vector on that surface, and ${\boldsymbol {\sf I}}$ is the identity tensor.

The dependence of the surface tension $\gamma$ on the surface concentration $\Gamma$ is given by the Langmuir equation of state \citep{T97}
\begin{equation}
\label{lan}
\gamma=\gamma_c+\Gamma_{\infty} R_g T\ln \left(1-\frac{\Gamma}{\Gamma_{\infty}}\right),  
\end{equation}
where $R_g$ is the gas constant and $T$ is the temperature. 

We assume that the liquid bath is not perturbed by the bubble in the upstream region $r=R_o$ and $z>0$ (Fig.\ \ref{sketch}). Therefore,
\begin{equation}
v_r=v_{\theta}=0,\quad v_z=-\frac{dh}{dt},\quad p+\rho g z=\text{const.}
\label{entranceflow}
\end{equation}
in that boundary. The non-reflecting boundary conditions 
\begin{equation}
\frac{\partial v_r}{\partial z}=\frac{\partial v_{\theta}}{\partial z}=\frac{\partial v_z}{\partial z}=0
\label{outflow}
\end{equation}
are applied in the downstream region far from the bubble ($r=R_o$ and $z<0$) to capture the wake (Fig.\ \ref{sketch}). The surfactant concentration at $r=R_o$ is $c_{\infty}$.

We consider the regularity conditions at the base flow symmetry axis $r=0$. The condition $v_z=0$ at the interface upper point allows us to calculate the bubble's vertical velocity in the steady base flow. We specify the bubble's volume in the base flow through the equation
\begin{equation}
V=\pi\int_0^{1} F^2\frac{\partial G}{\partial s}\, ds.
\end{equation}
Finally, we consider the following equation 
\begin{equation}
\frac{\partial F}{\partial s}\frac{\partial^2 F}{\partial s^2}+\frac{\partial G}{\partial s}\frac{\partial^2 G}{\partial s^2}=0.
\end{equation}
to ensure a uniform distribution of grid points along $s$.

The global linear stability of the steady solution is determined by calculating the eigenmodes. To this end, we assume the temporal dependence
\begin{equation}
\label{ss10}
\Psi=\Psi_0+\delta\Psi\, e^{-i\omega t+i m \theta}+\text{c.c.},  
\end{equation}
\begin{equation}
\label{ss2}
\Gamma=\Gamma_0+\delta\Gamma\, e^{-i\omega t+i m \theta}+\text{c.c.},  
\end{equation}
\begin{eqnarray}
\label{ss1}(r_i,z_i)=(r_{i0},z_{i0})+(\delta r_i,\delta z_i)\, e^{-i\omega t+i m \theta}+\text{c.c.},
\end{eqnarray}
where $\Psi(r,\theta,z;t)$ represents the unknowns $\{\mathbf{v}(r,\theta,z;t)$, $p(r,\theta,z;t)$, $c(r,\theta,z;t)\}$, and $\Psi_0(r,z)$ and $\delta \Psi(r,z)$ stand for the corresponding base flow (steady) solution and the eigenmode spatial dependence of the eigenmode, respectively. In addition, $\Gamma_0(s)$ and $\delta\Gamma(s)$ are surface coverage in the base flow and the eigenmode spatial dependence, respectively, whilst $(r_i,z_i)$ denotes the interface position, $(r_{i0},z_{i0})$ denotes the interface position in the base flow, and $(\delta r_i,\delta z_i)$ is the perturbation. In the global linear stability, one assumes that $|\delta \Psi|\ll |\Psi|$, $|\delta\Gamma|\ll \Gamma$, $|\delta r_i|\ll r_i$, and $|\delta z_i|\ll |z_i|$. Finally, $\omega=\omega_r+i\omega_i$ is the eigenfrequency characterizing the perturbation evolution of azimuthal number $m$. If the growth rate of the dominant mode (i.e., that with the largest $\omega_i$) is positive, then the base flow is asymptotically unstable under small-amplitude perturbations \citep{T11}. The flow is stable otherwise. Modes $m=\pm 1$ are the most unstable, so we focus on them for the remainder.

Eigenvalues and eigenfunctions are found by solving the generalized eigenvalue problem
\begin{equation}
\label{ei}
{\cal J}_b^{(p,q)}\delta \Psi^{(q)}=i\omega\, {\cal Q}_b^{(p,q)}\delta \Psi^{(q)}
\end{equation}
We calculate analytical expressions for the Jacobians ${\cal J}_b^{(p,q)}$ and ${\cal Q}_b^{(p,q)}$. If $\Psi$ is identified as $v_z$, $v_r$, and $imv_{\theta}$, Eq.\ (\ref{ei}) depends only on $m^2$, and matrices are real. Thus, eigenvalues and eigenfunctions appear in complex conjugate pairs $\omega=\pm\omega_r+i\omega_i$ and are the same for $m=\pm 1$.

The equations are numerically integrated with a variant of the boundary-fitted spectral method proposed by \citet{HM16a} \citep{JAM}. All the derivatives appearing in the governing equations are expressed in terms of the spatial coordinates ($\tilde{s}$,$\eta$) resulting from the mapping onto a rectangular domain using non-singular mapping. These equations are discretized in the $\eta$-direction with $n_{\eta}$ Chebyshev spectral collocation points to accumulate the grid points next to the interface \citep{HM16a}, facilitating the resolution of the surfactant boundary layer. We use second-order finite differences with $n_{s}$ equally spaced points to discretize the $\tilde{s}$-direction. The results presented in this work are calculated with $n_{\eta}=81$ and $n_{s}=451$. We conducted a grid sensitivity analysis to verify that the eigenvalue of the dominant mode was not significantly affected by a grid refinement (see Supplemental Material). The elements of the Jacobian corresponding to the linear system of equations are symbolic functions calculated by the symbolic package of Matlab before running the simulation. The eigenvalues are numerically found with the MATLAB function {\sc Eigs}.

\subsection{Experimental method}
\label{sec3}

This section describes the main aspects of the experimental method. More details can be found in the Supplemental Material. In an experiment, nitrogen was injected through a needle to form a bubble in the center of the water ($\rho=998$ kg/m$^3$ and $\mu=1.0016\times 10^{-3}$ kg/(m$\cdot$s)) tank bottom. We used needles with different diameters (30 $\mu$m, 50 $\mu$m, 62 $\mu$m, 75 $\mu$m, and 100 $\mu$m) to produce bubbles with radii $R=0.57$ mm, 0.66 mm, 0.70 mm, 0.76 mm, and 0.86 mm. After several seconds, the bubble detached from the needle and rose across the tank until it reached the free surface. We used a virtual binocular stereo vision system \citep{LWZGZXLL22} to image two perpendicular views of the rising bubble. The images were processed at the pixel level \citep{C86} to determine the bubble shape and velocity. This experimental method was validated by \citet{RVCMLH24} from comparison with the results for clean water obtained by \citet{D95}. 

\begin{figure}
\vspace{0.cm}
\begin{center}
\resizebox{0.45\textwidth}{!}{\includegraphics{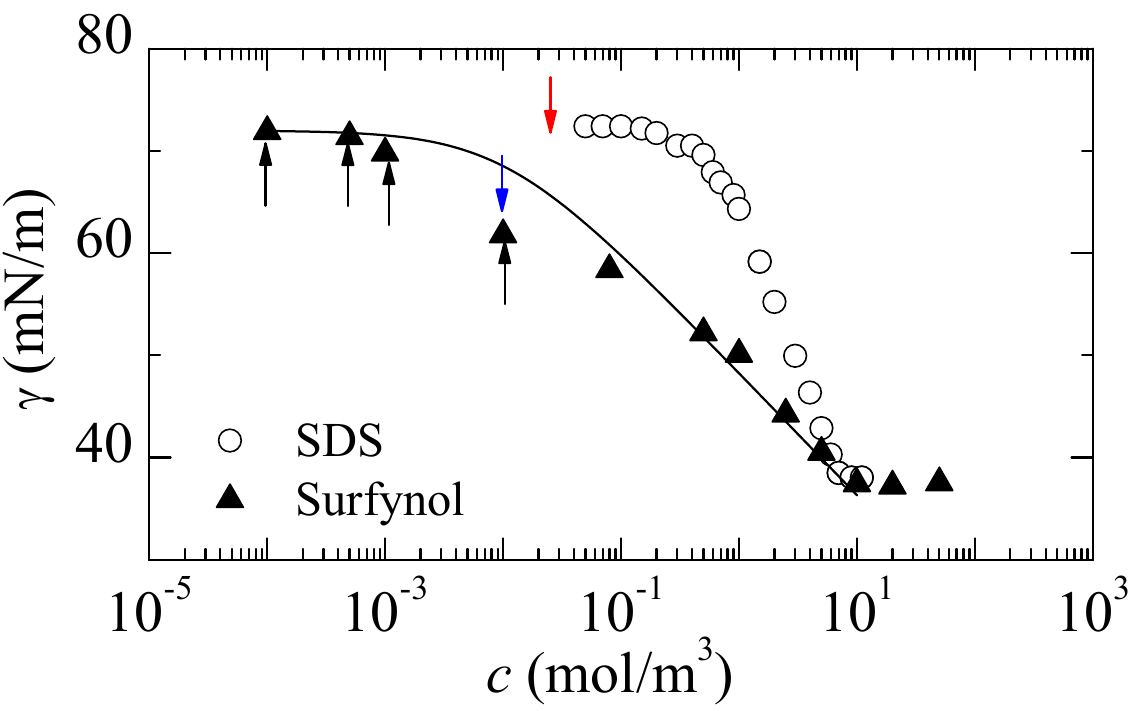}}
\end{center}
\caption{Surface tension $\gamma$ versus surfactant volumetric concentration $c$ for Surfynol\ 465 \citep{PZPA16} and SDS \citep{TMS70}. The line is the fit of the Langmuir equation of state $\gamma=\gamma_c-\Gamma_{\infty} R_g T\log(1+L_d\, c/\Gamma_{\infty})$ to the experimental data of Surfynol. The black arrows indicate the concentrations in our experiments with Surfynol. The blue arrow indicates the concentration in the global stability analysis (Sec.\ \ref{sec5}). The red arrow approximately indicates the maximum concentration considered in the analysis of \citet{RVCMLH24}.}
\label{st}
\end{figure}

We considered Surfynol\circledR\, 465 in our experiments. Surfynol is a non-ionic dimeric surfactant with two amphiphile parts (similar to monomeric surfactant molecules) connected by a bridge or a spacer \citep{PZPA16}. The experimental results were compared with those obtained with SDS \citep{TMS70}. Figure \ref{st} shows the dependence of the surface tension on the surfactant volumetric concentration in both cases. The Langmuir equation of state $\gamma=\gamma_c-\Gamma_{\infty} R_g T\log(1+L_d\, c/\Gamma_{\infty})$ was fit to the experimental data of Surfynol to calculate the value of $L_d$. \citet{PZPA16} accurately measured the value of $\Gamma_{\infty}$. We considered that value in our simulations. Table \ref{prop} shows the properties of SDS and Surfynol. 

\begin{table*}
\begin{center}
\begin{tabular}{ccccccc}
\hline
    &CMC  (mol/m$^3$) & $L_d$ (mm) & $(k_d)^{-1}$ & $\Gamma_\infty$ ($\mu$mol/m$^2$) & ${\cal D}$ (m$^2$/s) & ${\cal D}_S$ (m$^2$/s)\\ \hline
SDS & 8.0 & 0.037 & 0.16 & 3.2 & $8.0 \times 10^{-10}$ & $8.0 \times 10^{-10}$ \\ \hline
Surfynol 465 & 10 & 0.36 & -- & 2.1 & -- & -- \\ \hline
\end{tabular}
\end{center}
\caption{Properties of SDS and Surfynol.}
\label{prop}
\end{table*}

The surface tension for concentrations above the critical micelle concentration is practically the same in the two cases (Fig.\ \ref{st}). However, Surfynol is stronger than SDS at low concentrations. In this regime, the Langmuir equation of state (\ref{lan}) reduces to $\gamma_c-\gamma\simeq R_g\, T\, L_d\, c$. The values of the depletion length $L_d$ for both surfactants (Table \ref{prop}) indicate the larger strength of Surfynol in the dilute regime. It is worth noting that the maximum packing density $\Gamma_{\infty}$ is commonly used to characterize the surfactant strength. Interestingly, $\Gamma_{\infty}$ for SDS is larger than for Surfynol (Table \ref{prop}) despite Surfynol being much stronger than SDS for the low concentrations considered in this work. In fact, $\Gamma_{\infty}$ has little influence on the surfactant behavior in most of our experiments. 

A natural question is whether the effects of two surfactants must be compared considering the same absolute concentration $c$ or its relative value $c/c_{\text{cmc}}$. The answer probably depends on the purpose of the analysis. In this case, the critical micelle concentration of the two surfactants takes similar values; therefore, this question is irrelevant. Section \ref{sec4} compares the effects of Surfynol and SDS by considering the same ratio $c/c_{\text{cmc}}$. 

The Surfynol sorption constants have not been measured yet. In fact, a specific method must probably be developed for this purpose, given the very small sorption characteristic time of this surfactant for concentrations of the order of the critical micelle concentration \citep{VSQCC24}. It must be pointed out that the surfactant adsorption rate cannot be inferred from the equilibrium isotherm $\Gamma(c)$ (the depletion length $L_d$). For instance, the depletion length of Triton X-100 is much larger than that of SDS, even though its adsorption rate is lower than that of SDS {\em for the same value of} $c/c_{\text{cmc}}$ \citep{FCLHM25}. 

The adsorption rate must be inferred from a dynamic surface tension experiment. Using the maximum bubble-pressure tensiometry, \citet{VSQCC24} showed that the adsorption rate of Surfynol is much higher than that of SDS at the same relative concentration $c/c_{\text{cmc}}=1.3$. It is natural to hypothesize that the same occurs at lower concentrations. In this sense, Surfynol is a fast surfactant. 

The bulk and surface diffusion coefficients of Surfynol have not been determined experimentally either. One expects these coefficients to take values in the range $10^{-9}-10^{-10}$ m$^2$/s, as occurs to most surfactants with similar molecular weights \citep{T97}. We have verified that the terminal bubble velocity obtained in our simulation becomes independent of the diffusion coefficient for ${\cal D}\lesssim 10^{-7}$ m$^2$/s and is practically the same for ${\cal D}_S=10^{-6}$ m$^2$/s and ${\cal D}_S=5\times 10^{-7}$ m$^2$/s (see the Supplemental Material). The results presented in Sec.\ \ref{sec5} were calculated for ${\cal D}=10^{-8}$ m$^2$/s for ${\cal D}_S=5\times10^{-7}$ m$^2$/s. 

\begin{figure}
\vspace{0.cm}
\begin{center}
\resizebox{0.4\textwidth}{!}{\includegraphics{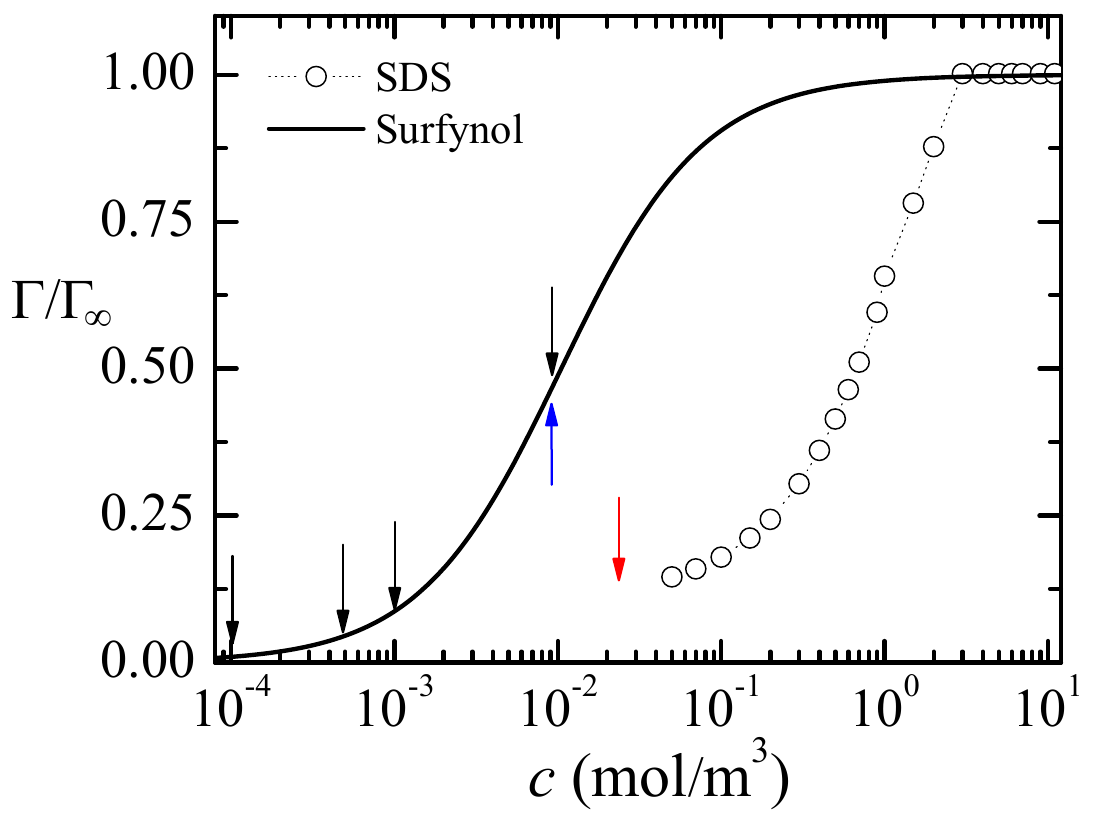}}
\end{center}
\caption{Surface coverage $\Gamma/\Gamma_{\infty}$ versus surfactant concentration $c$. The black arrows indicate the concentrations in our experiments with Surfynol. The blue arrow indicates the concentration in the global stability analysis (Sec.\ \ref{sec5}). The red arrow approximately indicates the maximum concentration considered in the analysis of \citet{RVCMLH24}.}
\label{st3}
\end{figure}

Figure \ref{st3} shows the dependence of the surfactant surface coverage $\Gamma/\Gamma_{\infty}$ on the volumetric concentration $c$. The values for SDS were measured by \citet{TMS70}, while the Surfynol results were obtained from the Langmuir isotherm (\ref{km}):
\begin{equation}
\frac{\Gamma}{\Gamma_{\infty}}=\frac{L_d\, c}{\Gamma_{\infty}+L_d\, c}.
\end{equation}
At low concentrations, this equation reduces to the Henry law $\Gamma=L_d\, c$, which allows one to understand the meaning of the term ``depletion length": $L_d$ is the distance perpendicular to the interface to contain the number of molecules adsorbed on the interface. As observed in Fig.\ \ref{st3}, the equilibrium surface coverage of Surfynol is much higher than that of SDS for the same volumetric concentration.

\begin{figure}
\vspace{0.cm}
\begin{center}
\resizebox{0.4\textwidth}{!}{\includegraphics{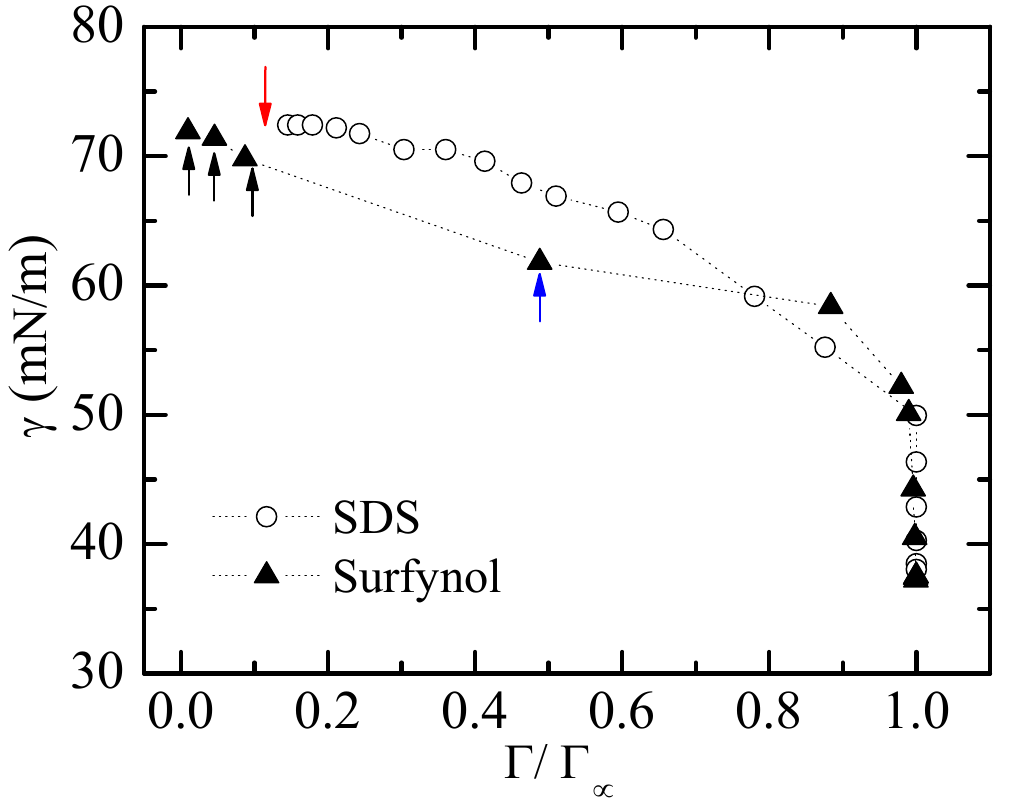}}
\end{center}
\caption{Surface tension $\gamma$ versus surfactant surface coverage $\Gamma/\Gamma_{\infty}$. The values for SDS were measured by \citet{TMS70}, while the surface tension $\gamma(\Gamma)$ for Surfynol was obtained from $\gamma(c)$ considering the relationship $\Gamma(c)$ given by the Langmuir isotherm (\ref{km}). The black arrows indicate the equilibrium surface coverages corresponding to our experiments with Surfynol. The blue arrow indicates the concentration in the global stability analysis (Sec.\ \ref{sec5}). The red arrow approximately indicates the maximum concentration considered in the analysis of \citet{RVCMLH24}.}
\label{st2}
\end{figure}

Figure \ref{st2} shows the dependence of the surface tension $\gamma$ on the surfactant surface coverage $\Gamma/\Gamma_{\infty}$. As mentioned above, Surfynol is much stronger than SDS for low surface coverages. Therefore, the Marangoni stress in our experiments with Surfynol can take similar values to those with SDS for much smaller Surfynol surface concentration gradients. This means that Surfynol is expected to be more smoothly distributed over the interface than SDS for the same value of $c/c_{\text{cmc}}$, facilitating the global stability analysis. 

\section{Dimensional Analysis}
\label{dim}

Both dimensional and dimensionless parameters provide useful information in this problem. This paper focuses on the surfactant effect at relatively low concentrations, for which the liquid surface tension is similar to that of water. Therefore, $R$, $\rho$, $\mu$, and $\gamma_c$ can be regarded as characteristic quantities in the presence of the surfactant. As mentioned in Sec.\ \ref{sec21}, the gas dynamics inside the bubble have negligible effects, allowing us to eliminate the gas density and viscosity from the analysis.

The following dimensionless numbers can be defined based on the characteristic quantities: the Bond number $B=\rho g R^2/\gamma_c$ and the Galilei number $\text{Ga}=\rho g^{1/2} R^{3/2}/\mu$. The effect of the monolayer is quantified through the dimensionless concentration $\hat{c}=c_{\infty}/c_{\text{cmc}}$ ($c_{\text{cmc}}$ is the critical micelle concentration) and the set of dimensionless parameters $\{{\cal P}_i\}$ characterizing the surfactant, as explained below. 

We consider the following dimensionless dependent parameters: the Reynolds number $\text{Re}=\rho v_t R/\mu$, where $v_t$ is the terminal velocity (the vertical velocity in the base flow), the aspect ratio $\chi=a/b$, where $a$ and $b$ are the half-length and half-breadth of the cross-sectional shape, respectively, and ${\cal S}=S/R^2$, where $S$ is the cross-sectional area. For a given gas-liquid-surfactant system, $B={\cal C}\, \text{Ga}^{4/3}$ and, therefore,
\begin{equation}
\label{dim1}
\text{Re}=\text{Re}(\text{Ga},\hat{c};\{{\cal P}_i\}),
\end{equation}
\begin{equation}
\label{dim2}
\chi=\chi(\text{Re},\hat{c};\{{\cal P}_i\}), \quad {\cal S}={\cal S}(\text{Re},\hat{c};\{{\cal P}_i\}),
\end{equation}
where we have considered that $\text{Ga}=\text{Ga}(\text{Re},\hat{c};\{{\cal P}_i\})$ in Eqs.\ (\ref{dim2}).

Consider the drag force $F_D$ experienced by the bubble in a steady vertical motion. The dimensionless force ${\cal F}_D=F_D/(\rho g R^3)$ is ${\cal F}_D=4\pi/3$. We define the drag coefficient $C_D=C_D(\text{Re},\hat{c};\{{\cal P}_i\})$ as
\begin{equation}
C_D=\frac{F_D}{1/2\, \rho v_t^2 S}=\frac{2{\cal F}_D}{\text{Ga}^{-2}\, \text{Re}^2 {\cal S}}.
\end{equation}

The following dimensionless parameters must be considered to characterize the physical properties of the fast surfactant monolayer: the bulk and surface Peclet numbers, $\text{Pe}=\ell_c v_c/{\cal D}$ and $\text{Pe}_s=\ell_c v_c/{\cal D}_S$, the dimensionless depletion length $\Lambda_d=L_d/\ell_c$, and the Marangoni (elasticity) number Ma=$\Gamma_{\infty} R_g T/\gamma_c$. In these expressions, $\ell_c=\mu^2/(\rho\gamma_c$) and $v_c=\gamma_c/\mu$ are the (intrinsic) viscous-capillary length and velocity, respectively. 

The dimensionless numbers mentioned above take fixed values for a given liquid-surfactant system. A fast surfactant is characterized by a relatively large value of $\Lambda_d$ and a relatively large adsorption rate. In this case, the sorption kinetics is characterized by the single parameter $\Lambda_d$ (see Sec.\ \ref{sec2}). Table \ref{t2ñ} shows the values of the intrinsic dimensionless numbers characterizing the surfactant.

\begin{table}
\begin{center}
    \begin{tabular}{ccccc}
    \hline
        & $\Lambda_d$ & Ma & $\text{Pe}$ & $\text{Pe}_s$\\
     \hline
    SDS  & $2.6\times 10^3$ & 0.11 & $1.3 \times 10^{3}$ & $1.3 \times 10^{3}$ \\
    \hline
    Surfynol & $2.6 \times 10^{4}$ & 0.071 & -- & -- \\
    \hline
    \end{tabular}
\end{center}
\caption{Values of the dimensionless numbers involving the physical properties of the fluids and surfactant monolayer.}
\label{t2ñ}
\end{table}

\section{Experimental results}
\label{sec4}

\subsection{Types of paths}

\begin{figure*}
\vspace{0.cm}
\begin{center}
\resizebox{1\textwidth}{!}{\includegraphics{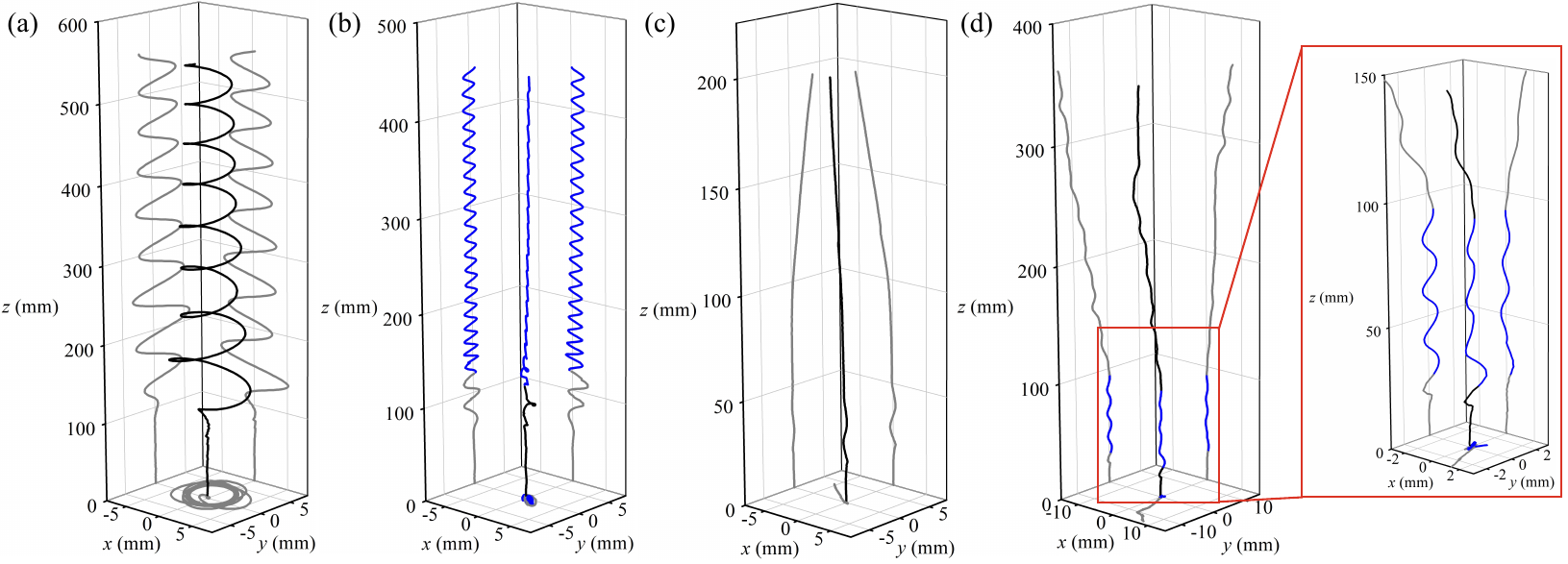}}
\end{center}
\caption{Bubble trajectory for $\text{Ga}=79$ and $c_{\infty}/c_{\text{cmc}}=10^{-5}$ (a), $\text{Ga}=79$ and $c_{\infty}/c_{\text{cmc}}=10^{-4}$ (b), $\text{Ga}=58$ and $c_{\infty}/c_{\text{cmc}}=10^{-3}$ (c), and $\text{Ga}=66$ and $c_{\infty}/c_{\text{cmc}}=10^{-3}$ (d). The graphs also show the projections of the trajectories onto the planes $(x,y)$, $(x,z)$, and $(y,z)$.}
\label{path}
\end{figure*}

\begin{figure*}
\vspace{0.cm}
\begin{center}
\includegraphics[width=0.49\textwidth]{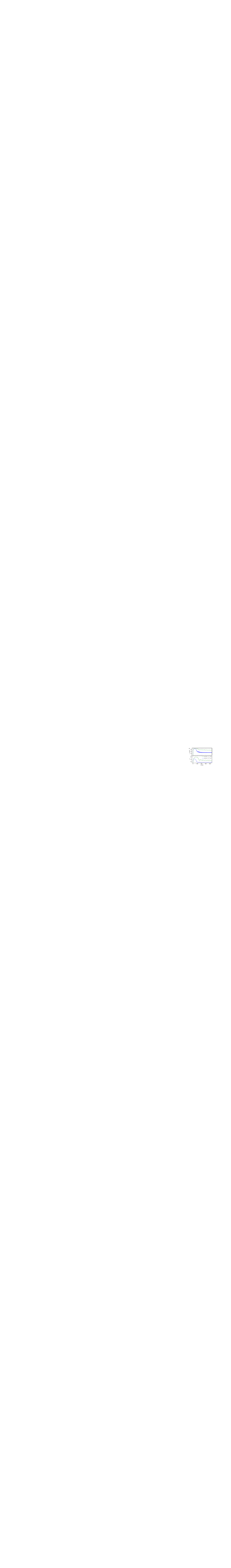}\includegraphics[width=0.51\textwidth]{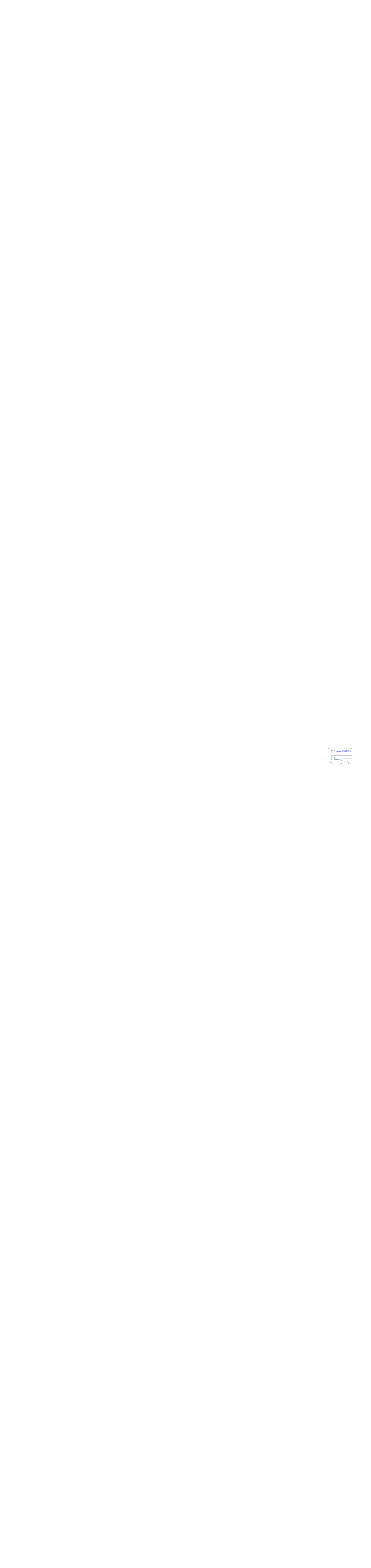}
\end{center}
\caption{Bubble vertical velocity $v_z$ and aspect ratio $\chi$ as a function of the vertical coordinate $z$ for the cases in Fig.\ \ref{path}.}
\label{pathz}
\end{figure*}

This section compares our experimental results for Surfynol with those obtained by \citet{RVCMLH24} for SDS. For illustration, Fig.\ \ref{path} shows four examples of the paths followed by bubbles in the presence of Surfynol. The corresponding bubble vertical velocity $v_z$ and aspect ratio $\chi$ are plotted in Fig.\ \ref{pathz} as a function of the vertical coordinate $z$.

Figure \ref{path}a corresponds to a helical motion characterized by a double-threaded wake (two counter-rotating vorticities) \citep{ERFM12}. Case (b) shows a zigzag motion obtained when the surfactant concentration increases for the same Galilei number (bubble radius). A vortex is shed periodically from the wake during the bubble rising \citep{ERFM12}. Oscillatory instabilities reduce the mean vertical velocity because energy transfers from the base flow to the unstable mode. This is observed in cases (a) and (b), in which the velocity decreases at $z\simeq 110$ mm due to the growth of the helical and zigzag motions (Fig.\ \ref{pathz}).

Figure \ref{path}c shows an oblique path with a tilt angle approximately equal to 2.5$^{\circ}$. This path results from the constant lift force generated by a pair of semi-infinite counter-rotating streamwise vortices (no vortex shedding occurs in this case) \citep{TTM14}. Oblique paths are a characteristic effect of the surfactant monolayer. They have been reported neither in experiments nor in simulations with clean water \citep{BSFM24}. These paths typically occur when the surfactant monolayer immobilizes the bubble surface \citep{TTM14}, which resembles the path instability of light spheres \citep{JBD03,JDB04}. One expects this instability to set in when both a sufficient vertical velocity has been reached and a sufficiently long wake has developed \citep{JDB04}. In the experiment of Fig.\ \ref{path}c, this occurs at a distance from the ejector $z\gtrsim 100$ mm, much larger than that at which the terminal velocity is reached ($z\simeq 25$ mm) (Fig.\ \ref{pathz}b). A significant decrease in the aspect ratio accompanies the growth of the stationary instability. The tiny oscillation observed for $z\lesssim 100$ mm disappears as the stationary instability grows. The initial conditions select the symmetry plane. 

Finally, Fig.\ \ref{path}d shows the transition from a zigzag to an oscillatory oblique path. The path tilt partially suppresses the bubble oscillations, slightly increasing the vertical velocity (Fig.\ \ref{pathz}b). The discrete Fourier transform of $v_z(t)$ exhibits a peak at the frequency $\omega\simeq 15$ Hz. As shown in Sec.\ \ref{sec51}, this frequency approximately corresponds to that of the unstable oscillatory mode in the secondary instability obtained in the global stability analysis. Oscillatory oblique paths are also observed when a light solid sphere rises in water \citep{ERFM12}.

The local velocity profiles exhibit the so-called ``overshooting"\ phenomenon: the velocity reaches a maximum and subsequently decreases to its terminal value (Fig.\ \ref{pathz}). The non-monotonous behavior of the bubble aspect ratio accompanies this phenomenon. Axisymmetric transient simulations have allowed us to explain the overshooting with non-fast surfactants in terms of the extra drag resulting from the late growth of the dynamic surfactant layer \citep{FCLHM25}. Interestingly, the same phenomenon occurs with Surfynol despite its much faster sorption kinetics. This suggests that surfactant diffusion prevents the full growth of the dynamic surfactant layer during the bubble acceleration.

\begin{figure}[h]
\begin{center}
\resizebox{0.45\textwidth}{!}{\includegraphics{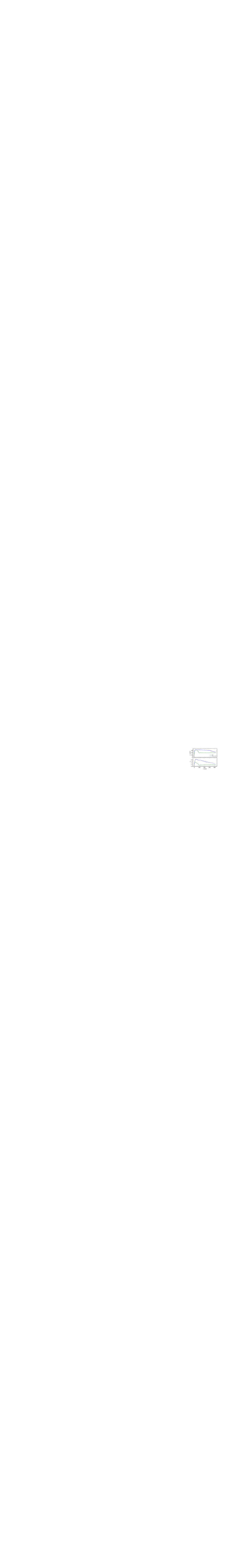}}
\end{center}
\caption{Bubble velocity $v_z$ and aspect ratio $\chi$ as a function of the vertical position $z$ of the center of gravity for $\text{Ga}=66$ and $c_{\infty}/c_{\mbox{\small cmc}}=10^{-3}$ \citep{FCLHM25}. The solid and dashed lines correspond to the stable and oscillatory parts of the bubble trajectory, respectively.} 
\label{duplicity2}
\end{figure}

Figure \ref{duplicity2} shows experimental results for SDS and Triton X-100 obtained by \citet{FCLHM25} for the same values $\text{Ga}=66$ and $c_{\infty}/c_{\text{cmc}}=10^{-3}$ as those in Fig. \ref{pathz}b. The time interval (length) over which the overshooting occurs increases as the desorption rate decreases. This explains why the overshooting extends over a much longer time with Triton. The comparison between these results and those of Surfynol indicates that Surfynol desorption takes place much faster (overshooting occurs for $z\lesssim 20$ mm in this case). Bubble rising can be used as a testbed to evaluate qualitatively the desorption rate of a fast surfactant.

\begin{figure*}
\begin{center}
\resizebox{0.65\textwidth}{!}{\includegraphics{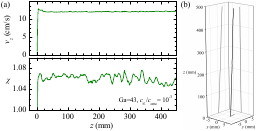}}
\end{center}
\caption{Bubble velocity $v_z$ and aspect ratio $\chi$ as a function of the vertical position $z$ of the center of gravity for $\text{Ga}=43$ and $c_{\infty}/c_{\mbox{\small cmc}}=10^{-3}$ (a) and its corresponding trayectory (b).} 
\label{duplicity3}
\end{figure*}

The experiment with $\text{Ga}=43$ and $c_{\infty}/c_{\text{cmc}}=10^{-3}$ exhibits a slightly oblique path (Fig.\ \ref{duplicity3}). The tilt angle is approximately 0.4$^{\circ}$, smaller than the threshold (0.5$^{\circ}$) adopted to categorize the path as oblique. Unlike in the case (c) of Fig.\ \ref{path}, the path is inclined right after the bubble ejection, and the aspect ratio does not significantly change during the bubble rising. As shown in Sec.\ \ref{sec51}, the global stability analysis predicts a stable path for these experimental conditions. The bubble radius ($R=0.57$ mm) is the smallest in our experiments. The bubble rising is the slowest, and the surfactant monolayer is practically formed when the maximum velocity is reached, despite the relatively large surfactant concentration $c_{\infty}/c_{\text{cmc}}=10^{-3}$. Consequently, the overshooting phenomenon is hardly observed (Fig.\ \ref{duplicity3}).

\subsection{The transition from clean to immobilized interface}

\begin{figure}
\vspace{0.cm}
\begin{center}
\resizebox{0.4\textwidth}{!}{\includegraphics{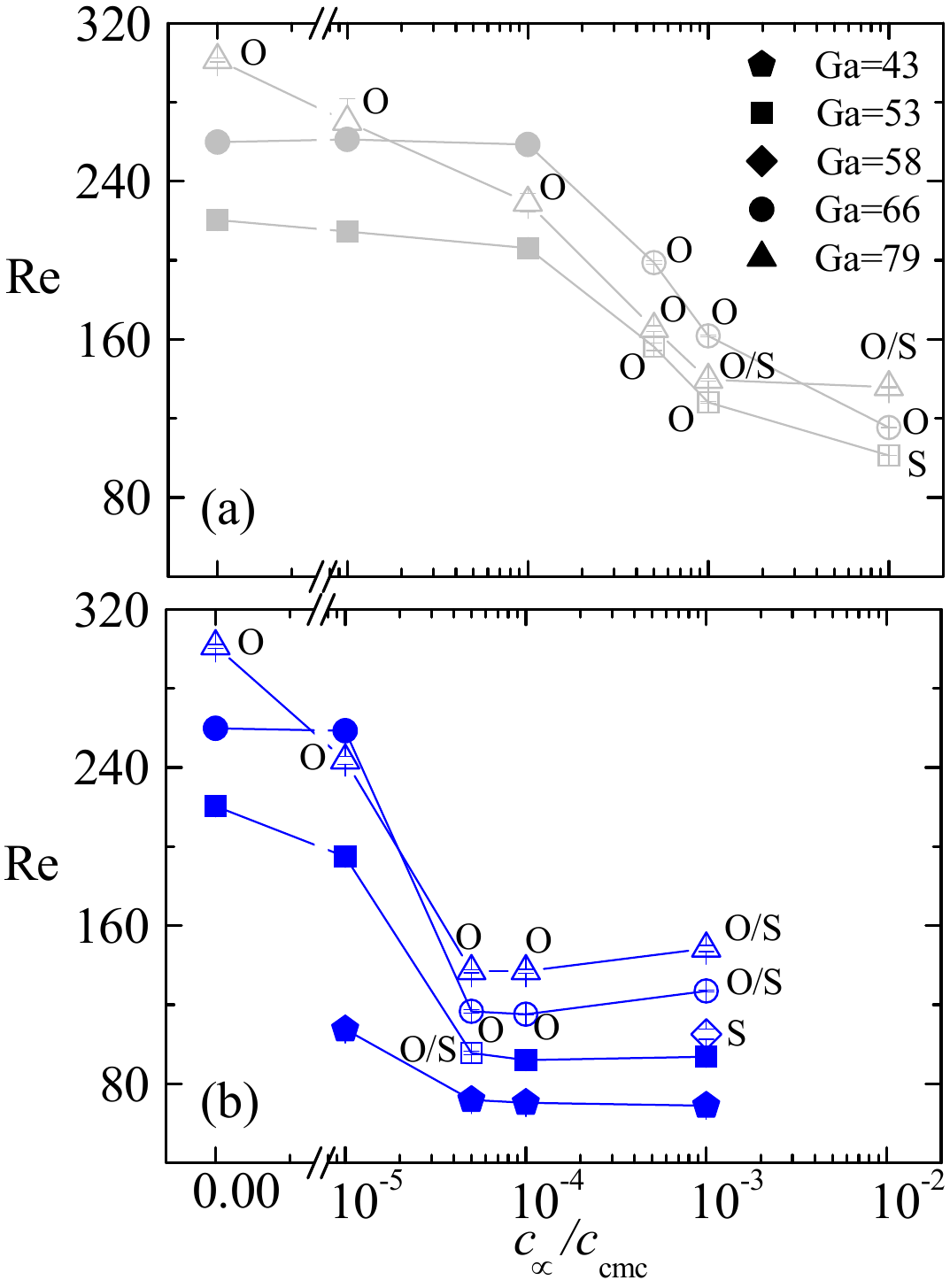}}
\end{center}
\caption{Reynolds number Re as a function of the surfactant concentration $c_{\infty}/c_{\text{cmc}}$ for SDS (a) and Surfynol (b). The solid and open symbols correspond to stable and unstable realizations, respectively. The labels S, O, and S/O indicate whether the stability is stationary, oscillatory, or a combination of both. The error bars in the SDS data correspond to the standard deviation and show the high reproducibility of our experimental method.}
\label{Re_c}
\end{figure}

Consider the Reynolds number (the terminal velocity) measured for a fixed value of the Galilei number (bubble radius) in the presence of SDS (Fig.\ \ref{Re_c}a). The surfactant produces small effects for low surfactant concentrations. In this regime, the Reynolds number is slightly smaller than that corresponding to the surfactant-free case. As the surfactant concentration increases, the Reynolds number decreases to a value approximately equal to that corresponding to a solid sphere. In this regime, the surfactant distributes smoothly along the interface, which is immobilized by the Marangoni stress \citep{RVCMLH24}. 

One can identify an intermediate regime corresponding to surfactant concentrations between those of nearly clean bubble and practically-immobilized interface. In this intermediate regime, SDS is absorbed in the bubble front and transported by convection toward the rear, where it is desorbed. Convection and surface diffusion compete within a short interface portion, forming a surfactant surface boundary layer \citep{RVCMLH24}. Only the interface behind that layer is immobilized (the stagnation cap model). The size of the immobilized interface portion increases with the surfactant concentration until it occupies the entire bubble surface (the immobilized interface regime).

In the case of Surfynol, the Reynolds numbers corresponding to the immobilized interface regime are reached at much smaller values of $c_{\infty}/c_{\text{cmc}}$ (Fig.\ \ref{Re_c}b). This can be explained in terms of the surfactant behavior at equilibrium. The Surfynol depletion length is much larger than that of SDS, which implies that more Surfynol is expected to be adsorbed during the bubble rising (Fig.\ \ref{st3}). Besides, Surfynol is stronger than SDS for small surface coverages (Fig.\ \ref{st2}). Both effects increase the Marangoni stress and, consequently, the drag force. 

\begin{figure}
\vspace{0.cm}
\begin{center}
\resizebox{0.4\textwidth}{!}{\includegraphics{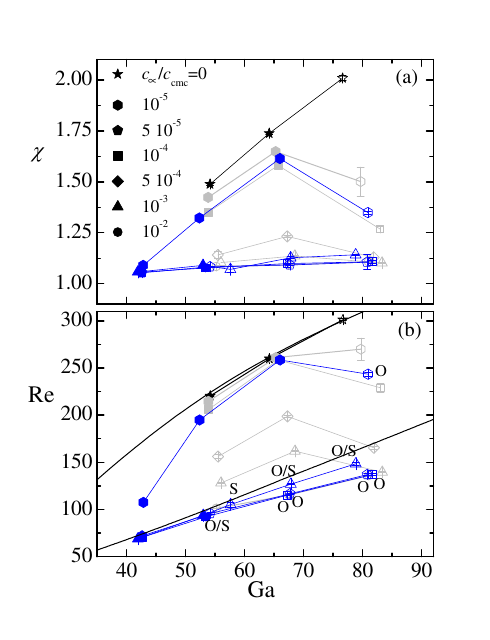}}
\end{center}
\caption{(a) Aspect ratio $\chi$ and (b) Reynolds number Re as a function of the Galilei number Ga for SDS (grey symbols) and Surfynol (blue symbols). The solid and open symbols correspond to stable and unstable realizations, respectively. The labels S, O, and S/O indicate whether the stability is stationary, oscillatory, or a combination of both. The error bars correspond to the standard deviation and show the high reproducibility of our experimental method. The upper and lower solid lines are the function Re(Ga) for a clean bubble \citep{HE23} and a solid hollow sphere, respectively.}
\label{ReGa}
\end{figure}

The transition from the clean bubble behavior to the immobilized interface regime occurs within a narrower interval of $c_{\infty}/c_{\text{cmc}}$ in the Surfynol case. For instance, consider the experiments with $\text{Ga}=66$. The Reynolds number decreases approximately from the clean surface value $\text{Re}=258.7$ at $c_{\infty}/c_{\text{cmc}}=10^{-5}$ to the immobilized interface value $\text{Re}=116.6$ at $c_{\infty}/c_{\text{cmc}}=5\times 10^{-4}$. Conversely, several experimental realizations with intermediate Reynolds numbers were found in the presence of SDS (Fig.\ \ref{ReGa}). This difference can be explained as follows. The Surfynol surface concentration is practically at equilibrium with the volumetric one at any interface point. Therefore, a surfactant surface boundary layer entails large gradients of the volumetric concentration next to the interface in the streamwise direction. However, strong surfactant convection parallel to the bubble surface reduces these gradients. Therefore, the surfactant surface boundary layer and stagnation caps, characteristic of the above-mentioned intermediate regime, appear under more stringent parameter conditions with fast surfactants.

Figure \ref{ReGa} also shows the difference between SDS and Surfynol discussed above. A sharp transition occurs at the Surfynol concentration $c_{\infty}/c_{\text{cmc}}\simeq 10^{-4}$. For $c_{\infty}/c_{\text{cmc}}\leq 10^{-4}$, the bubble shape significantly deviates from the sphere, and the Reynolds number values are relatively close to those of the clean bubble. Both $\chi$ and Re exhibit a non-monotonous dependency on Ga. For $c_{\infty}/c_{\text{cmc}}\geq 10^{-4}$, the bubble adopts a quasi-spherical shape ($\chi\lesssim 1.1$) and the Reynolds number sharply decreases. In this concentration regime, both $\chi$ and Re increase linearly with Ga. 

The upper and lower solid lines in Fig.\ \ref{ReGa}b are the function Re(Ga) for a clean bubble \citep{HE23} and a solid hollow sphere \citep{CGW78}, respectively. For Surfynol concentrations $c_{\infty}/c_{\text{cmc}}\geq 10^{-4}$, the Reynolds number falls below the solid line corresponding to a solid sphere. This can be explained by the drag force increase due to the small bubble deformation ($\chi\simeq 1.1$) and the velocity decrease caused by the oscillatory instability.

\begin{figure}
\vspace{0.cm}
\begin{center}
\resizebox{0.4\textwidth}{!}{\includegraphics{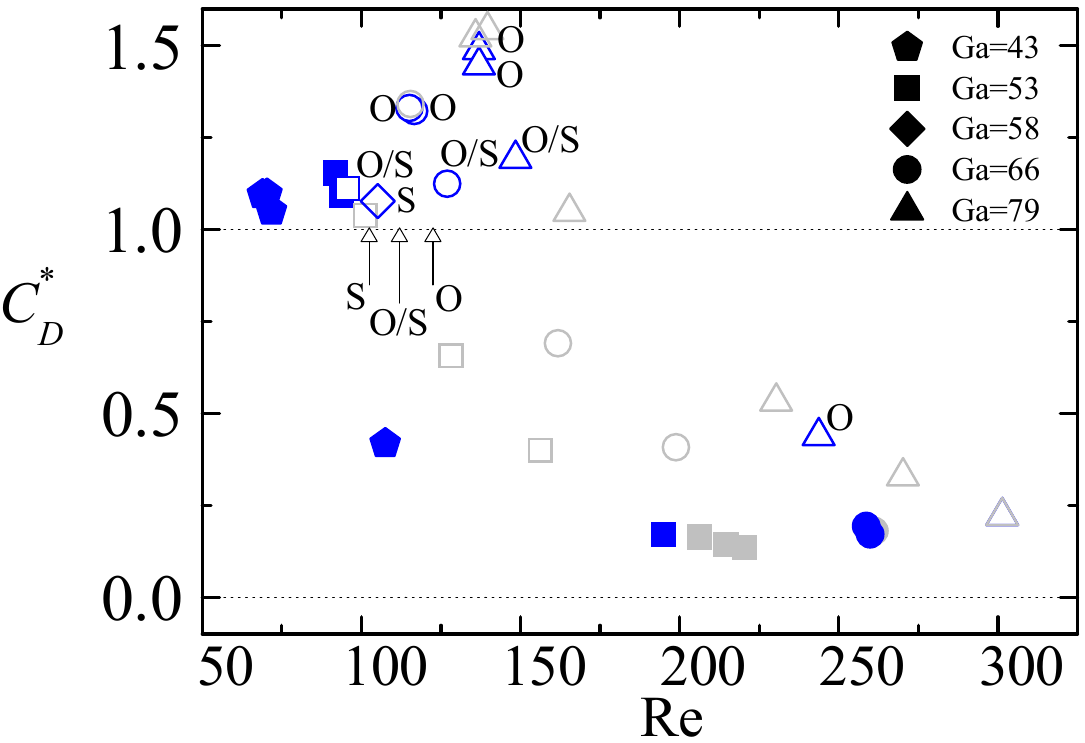}}
\end{center}
\caption{Normalized drag coefficient $C_D^*$ as a function of the Reynolds number Re. The solid and open symbols correspond to stable and unstable realizations, respectively. The labels S, O, and S/O indicate whether the stability is stationary, oscillatory, or a combination of both. The arrows indicate the critical Reynolds numbers mentioned in the text.}
\label{cdre}
\end{figure}

Figure \ref{cdre} shows the projection of the experimental data on the parameter plane (Re,$C_D^*$). Here, $C_D^*$ is the normalized drag coefficient defined as
\begin{equation}
C_D^*=\frac{C_D-C_{D,\text{clean}}}{C_{D,\text{rigid}}-C_{D,\text{clean}}},
\end{equation}
where $C_{D,\text{clean}}$ and $C_{D,\text{rigid}}$ are the drag coefficient of a spherical bubble with the free-slip condition (clean interface) \citep{MKL94} and no-slip condition (rigid interface) \citep{CGW78}, respectively. It must be noted that the effect of the bubble shape on $C_D$, $C_{D,\text{clean}}$, and $C_{D,\text{rigid}}$ is much smaller than the influence of the interface condition for the range of Reynolds numbers considered in our analysis \citep{TTM14}. Therefore, the coefficient $C_D^*$ essentially indicates the degree of slip on the bubble surface. The free-slip and no-slip conditions correspond to $C_D^*=0$ and 1, respectively.

Consider the experiments with Surfynol. For the higher surfactant concentrations, $C_D^*$ is significantly larger than unity. This effect is smaller without oscillations (solid symbols) and is explained by the less aerodynamic shape adopted by the bubble ($\chi>1$). The bubble oscillation reduces the terminal velocity, increasing the drag coefficient up to 1.5. The magnitude of this effect depends on the type and amplitude of the oscillation. The same phenomenon is observed with SDS. 

\subsection{Path instability transition}

Now, we focus on path instability. In Figs.\ \ref{Re_c}-\ref{cdre}, we distinguish between the oscillatory (O) instability leading to either helical or zigzag motion and the stationary (S) instability leading to an oblique path with a tilt angle greater than 0.5$^{\circ}$. There are several experimental realizations where the oscillatory and stationary instabilities coexist. These realizations are denoted with the symbol O/S. 

Significant differences exist between the path stability in the presence of SDS and Surfynol. In the case of SDS, oscillatory instability arises for Reynolds numbers larger than those in the immobilized interface regime. This does not occur in the Surfynol case, where unstable realizations can be found only at Reynolds numbers similar to that of the immobilized interface case (Fig.\ \ref{ReGa}). 

All the experimental realizations with relatively large concentrations ($c_{\infty}/c_{\text{cmc}}\geq 10^{-3}$) of SDS are unstable. Conversely, path instability arises at a critical Galilei number for the Surfynol concentration $c_{\infty}/c_{\text{cmc}}=10^{-3}$ (Fig.\ \ref{Re_c}). As explained in Sec.\ \ref{sec5}, this is a critical characteristic of Surfynol because it allows us to study the instability transition from the global stability analysis.

In the SDS case, the stability character always exhibits a monotonous dependence with respect to the surfactant concentration: for a fixed value of Ga (bubble radius), instability arises above a critical surfactant concentration. Interestingly, this does not occur with Surfynol for $\text{Ga}=53$ (Fig.\ \ref{Re_c}). In this case, the path becomes unstable at $c_{\infty}/c_{\text{cmc}}=5\times 10^{-5}$ and stability is recovered for larger concentrations. 

The critical Reynolds number for the path stability of a solid sphere is approximately 102.5 \citep{JDB04}. The instability corresponds to a non-oscillating growth of small non-axisymmetric perturbations, which gives rise to a stationary oblique path (S in our notation). A secondary Hopf bifurcation occurs at $\text{Re}=112$, leading to an oblique oscillatory path (O/S in our notation). Finally, a low-frequency zigzag (O in our notation) arises at $\text{Re}=122.5$ \citep{JDB04}. The above-mentioned critical Reynolds numbers are indicated by arrows at the line $C_D^*=1$ in Fig.\ \ref{cdre}. The same sequence of instabilities is observed with Surfynol for $c_{\infty}/c_{\text{cmc}}=10^{-3}$ and $\text{Ga}\geq 53$. The surfactant practically immobilizes the interface at this concentration, which explains why the sequence of instabilities is the same as that of a solid sphere \citep{JDB04}. We analyze this sequence of instabilities from global linear stability in Sec.\ \ref{sec5}.

\section{Global linear stability}
\label{sec5}

In this section, we determine the critical Galilei number (bubble radius) for a given surfactant concentration and elucidate the mechanism responsible for the instability. Hereafter, and unless otherwise stated, length, velocity, surfactant flux, and stress are measured in terms of $R$, $v_t$, ${\cal D}/L_d c_{\infty}$, and $\rho v_t^2$, respectively. At very small surfactant concentrations, the surfactant surface distribution results from the coupling between the extremely thin surfactant boundary layers near and over the interface. This considerably complicates the calculation of both the base flow and the linear eigenmodes. In fact, we have verified that the critical eigenmode is sensitive to the (unknown) value of the diffusion coefficient even for ${\cal D}<10^{-8}$ m$^2$/s. To circumvent this obstacle, we will consider only a moderately large surfactant concentration, $c_{\infty}/c_{\text{cmc}}=10^{-3}$, for which the surfactant distributes smoothly on the interface. 

\subsection{Validation of the numerical solution}
\label{sec51}

Table \ref{com} compares the terminal velocity, aspect ratio, and the stability character obtained from the global linear stability with their counterparts in the experiment. There is a good agreement for the three Galilei numbers. The numerical terminal velocities are slightly larger than the experimental values. The bifurcation sequence predicted by the global stability analysis is found in the experiments. The path is stable for $\text{Ga}=53$. A stationary instability arises at $\text{Ga}=58$. A secondary oscillatory Hopf bifurcation occurs at Ga$\simeq 66$. The stationary and oscillatory instabilities coexist at $\text{Ga}=66$. This corresponds to the paths observed in the experiments: a straight path for $\text{Ga}=53$, an oblique path for $\text{Ga}=58$, and oscillations around an oblique path for $\text{Ga}=66$. The oscillation frequency predicted by the linear stability analysis for $\text{Ga}=66$ is 22 Hz, while the corresponding experimental value was around 15 Hz. 

\begin{table*}
\begin{center}
\begin{tabular}{cccc}
\hline
Ga & 53 & 58 & 66 \\
\hline
$v_t$ (cm/s) (Exp. \& GSA) & 14.21 \& 14.4 & 15.08 \& 15.70 & 16.34 \& 16.97 \\
\hline
$\chi$ (Exp. \& GSA) & 1.09 \& 1.03 & 1.07 \& 1.04 & 1.13 \& 1.05 \\
\hline
Stability (Exp. \& GSA) & Stable \& Stable & S \& S & S/O \& S/O \\
\hline
\end{tabular}
\end{center}
\caption{Experimental and numerical results for different Galilei numbers around its critical value for $c_{\infty}/c_{\text{cmc}}=10^{-3}$. The table shows the terminal velocity $v_t$, aspect ratio $\chi$, and the stability character both in the experiment (Exp.) and obtained from the global stability analysis (GSA).}
\label{com}
\end{table*}

\subsection{Base flow}
\label{sec52}

We will return to the global stability analysis in Sec.\ \ref{sec53}. Now, we examine the base flow for $\text{Ga}=53$, 58, and 66. Consider the marginally stable flow obtained for the critical Galilei number $\text{Ga}=58$. Figure \ref{Streamlines} shows the flow streamlines and the volumetric surfactant concentration. As shown below, the surfactant monolayer practically immobilizes the interface. The viscous boundary layer remains attached to the bubble front and separates from the bubble surface behind the bubble equator. Two recirculation cells arise in the wake. The surfactant concentration in the bubble front is lower than the upstream value $c_{\infty}$. The concentration increases and reaches its maximum value approximately at the boundary layer separation point (indicated by a circle in Fig.\ \ref{Streamlines}). As shown below, the surfactant is adsorbed at the bubble front, while desorption occurs beyond a certain point in front of the bubble's equator. The surfactant excess is transferred back to the liquid by diffusion.

\begin{figure*}
\begin{center}
\resizebox{0.3\textwidth}{!}{\includegraphics{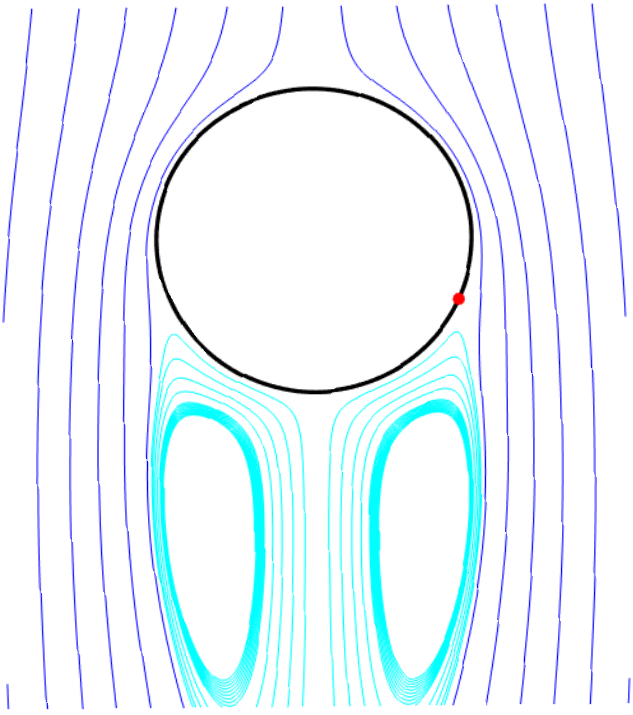}}
\resizebox{0.3\textwidth}{!}{\includegraphics{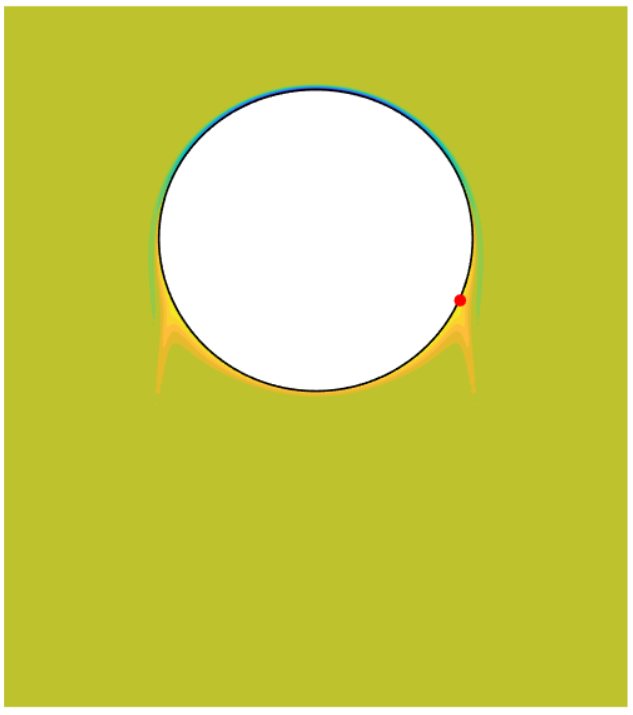}}
\resizebox{0.21\textwidth}{!}{\includegraphics{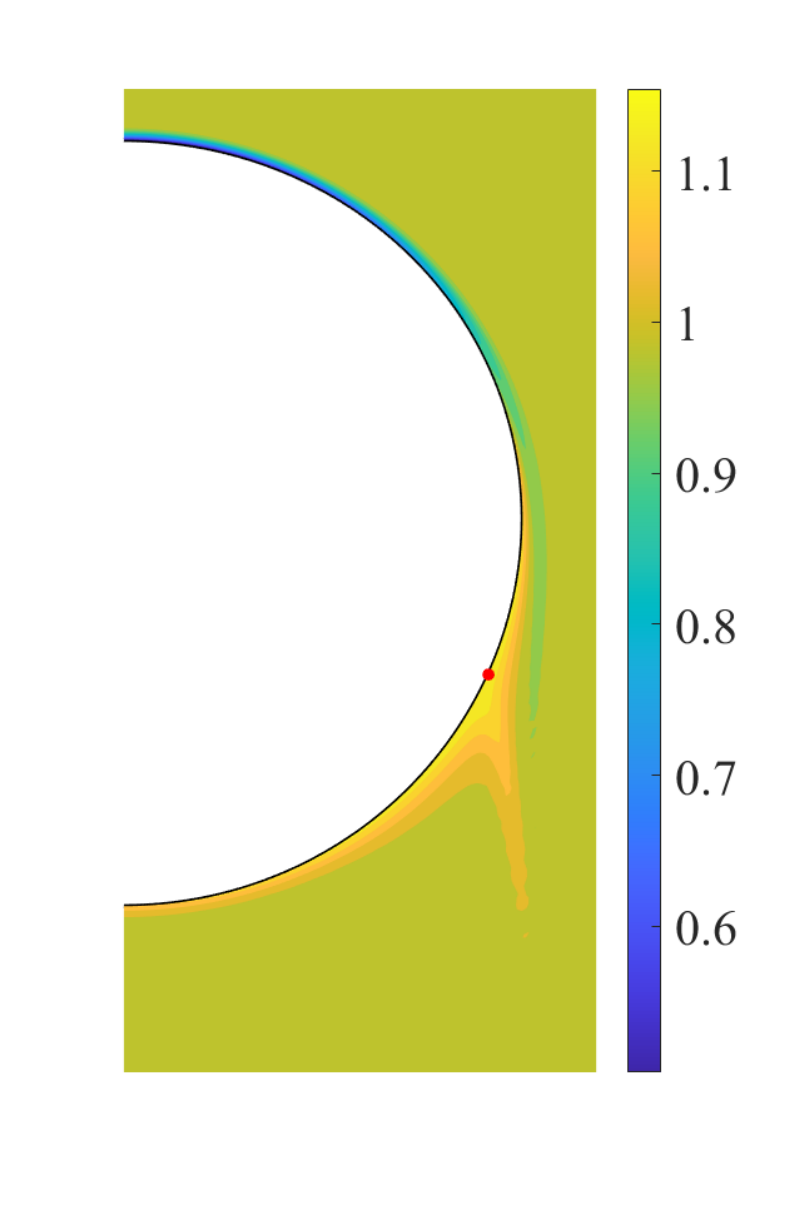}}
\end{center}
\caption{Streamlines (left) and surfactant volumetric concentration $c/c_{\infty}$ (right) for $c_{\infty}/c_{\text{cmc}}=10^{-3}$ and $\text{Ga}=58$. The red circle indicates the viscous boundary layer separation point.} 
\label{Streamlines}
\end{figure*}

\begin{figure}
\begin{center}
\resizebox{0.5\textwidth}{!}{\includegraphics{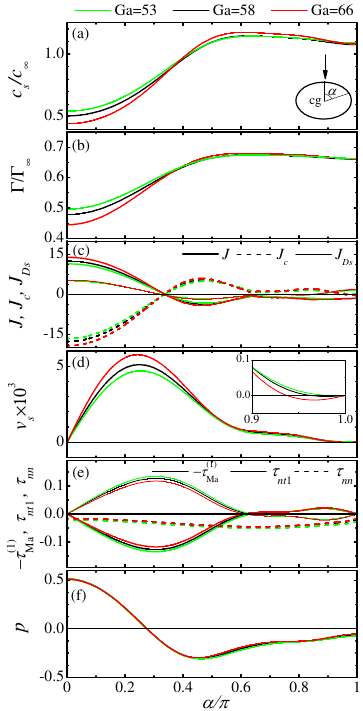}}
\end{center}
\caption{Volumetric surfactant concentration $c_s$ evaluated at the free surface (a), surfactant surface concentration $\Gamma$ (b), surfactant fluxes (c), surface velocity $v_s$ (d), Marangoni stress $\tau_{\text{Ma}}^{(1)}$, tangential viscous stress $\tau_{nt1}$, and normal viscous stress $\tau_{nn}$ (e), and pressure $p$ (f) as a function of the polar angle $\alpha$ for $c_{\infty}/c_{\text{cmc}}=10^{-3}$.} 
\label{profiles}
\end{figure}

Figure \ref{profiles} shows the distribution of the relevant quantities over the bubble surface. As the sublayer and the interface are at equilibrium, both the bulk (Fig.\ \ref{profiles}a) and the surface concentration (Fig. \ref{profiles}b) show the same behavior. Both concentrations show a minimum at the front stagnation point and increase smoothly downstream. Their maximum values are reached behind the bubble equator ($\alpha/\pi=0.63$), not at the rear of the bubble. This occurs due to the reverse flow in the wake, which convects the surfactant from the south pole to the equator. 

Figure \ref{profiles}c shows the values of the fluxes in the surfactant conservation equation (\ref{ad}); i.e., the flux of surfactant exchanged with the sublayer, $J={\cal D}\left. \nabla c\right|_n$, the flux convected over the interface, $J_c={\boldsymbol \nabla}_S\cdot(\Gamma {\bf v}_s)$, and the flux difussed across the interface, $J_{Ds}={\cal D}_S\nabla_s^2 \Gamma$. In all the cases, positive (negative) values imply that the surfactant is gained (lost) in the interface element. Due to uneven surfactant distribution, surface tension gradients cause Marangoni stresses that almost immobilize the interface. However, surfactant convection plays a crucial role. Convection pushes the surfactant away from the bubble front. Surface diffusion does not compensate for convection. Therefore, the surfactant is transferred from the sublayer in this part of the bubble. 

The velocity $v_s$ (Fig.\ \ref{profiles}d) and the surfactant gradient $d\Gamma/ds$ decrease in the bubble south hemisphere, resulting in a reduction of the convection flux magnitude in that region (Fig.\ \ref{profiles}c). This flux becomes positive, implying that convection accumulates surfactant in this part of the bubble. Most of the surfactant gained by convection is transferred back to the liquid. For this reason, the sublayer concentration grows downstream along the interface (Fig.\ \ref{Streamlines}-right), and the interface and sublayer remain at equilibrium.

Convection plays a key role in maintaining the variation of surfactant concentration over the interface, even though the surface velocity is three orders of magnitude smaller than the terminal velocity. The bulk and surface diffusion collaborate to reduce the surfactant gradient. Since the equilibrium equation (\ref{km}) links $c_s$ and $\Gamma$, both mechanisms transfer surfactant in the same direction over most of the interface. This explains why $J$ and $J_{Ds}$ have the same sign for most values of $\alpha$ (Fig.\ \ref{profiles}c). This also explains why $J$ vanishes near the inflection point $\nabla_s^2 \Gamma=0$ ($J_{Ds}=0$), implying that $J_c$ vanishes near that point too. 

The Marangoni stress almost immobilizes the interface. The tiny values taken by the surface velocity (Fig.\ \ref{profiles}d) are crucial to understanding the surfactant behavior. In the absence of surfactant convection (a fully immobilized interface), diffusion would render surfactant concentration uniform, eliminating the Marangoni stress responsible for interface immobilization. 

The Marangoni stress $\tau_{\text{Ma}}^{(1)}$ is balanced by the viscous tangential stress $\tau_{nt1}$ (Fig.\ \ref{profiles}e). The stress $\tau_{\text{Ma}}^{(1)}$ and, therefore, $\tau_{nt1}$ vanish where the surfactant surface concentration is maximum ($d\Gamma/ds=d\gamma/ds=0$). We conclude that the boundary layer separates from the bubble surface where the concentration is maximum. The sign of $\tau_{\text{Ma}}^{(1)}$ changes in the wake, opposing the reverse flow in that region. This prevents the interface from moving toward the bubble front except within a region very close to the bubble's south pole (see the inset in Fig.\ \ref{profiles}d). Both the low pressure in the wake (Fig. \ref{profiles}f) and the viscous stress contribute to the drag.

Figure \ref{profiles} also shows the results for a subcritical ($\text{Ga}=53$) and a supercritical ($\text{Ga}=65$) case. The bubble terminal velocity increases with Ga (the bubble radius) (Fig.\ \ref{profiles}c), which enhances surfactant convection (Fig.\ \ref{profiles}d). As mentioned above, surface and bulk diffusion compensate for the surfactant depletion caused by convection in the bubble front. Therefore, the magnitude of these two mechanisms also increases with Ga (Fig.\ \ref{profiles}d). This also explains why $c_s$ decreases at the front stagnant point as Ga increases (Fig.\ \ref{profiles}a).

The Marangoni stress in terms of the dynamical pressure $\rho v_t^2$ decreases as Ga increases (Fig.\ \ref{profiles}e), resulting in less interface immobilization (Fig. \ref{profiles}d). The variation of the Galilei number significantly affects neither the surfactant concentration inflection point ($J_{Ds}=0$ and $J\simeq J_c\simeq 0$) (Fig.\ \ref{profiles}c) nor the boundary layer separation point ($\tau_{\text{Ma}}^{(1)}=\tau_{nt1}=0$) (Fig.\ \ref{profiles}e). In terms of the dynamical pressure force $\rho v_t^2\, S$, drag decreases as Ga increases due to a decrease in viscous stress and a tiny increase in the pressure at the wake. The reverse flow near the bubble's south pole disappears for the subcritical case $\text{Ga}=53$ (see the inset in Fig.\ \ref{profiles}d).

Overall, the base flow for the subcritical, critical, and supercritical Galieli numbers hardly differ. The mechanisms responsible for the instability must be described by analyzing the effect of the perturbation associated with the critical eigenmode.

\subsection{Analysis of the instability}
\label{sec53}

This section analyzes the eigenmodes for both subcritical and supercritical Galieli numbers. For this purpose, the eigenfunctions are normalized by setting $\text{max}(\delta r_i)=1$. As mentioned above, the lengths, velocities, and stresses are divided by $R$, $v_t$, and $\rho v_t^2$, respectively. Consequently, the forces and torques are divided by $\rho v_t^2 R^2$ and $\rho v_t^2 R^3$, respectively.

\begin{figure}
\vspace{0.cm}
\begin{center}
\resizebox{0.45\textwidth}{!}{\includegraphics{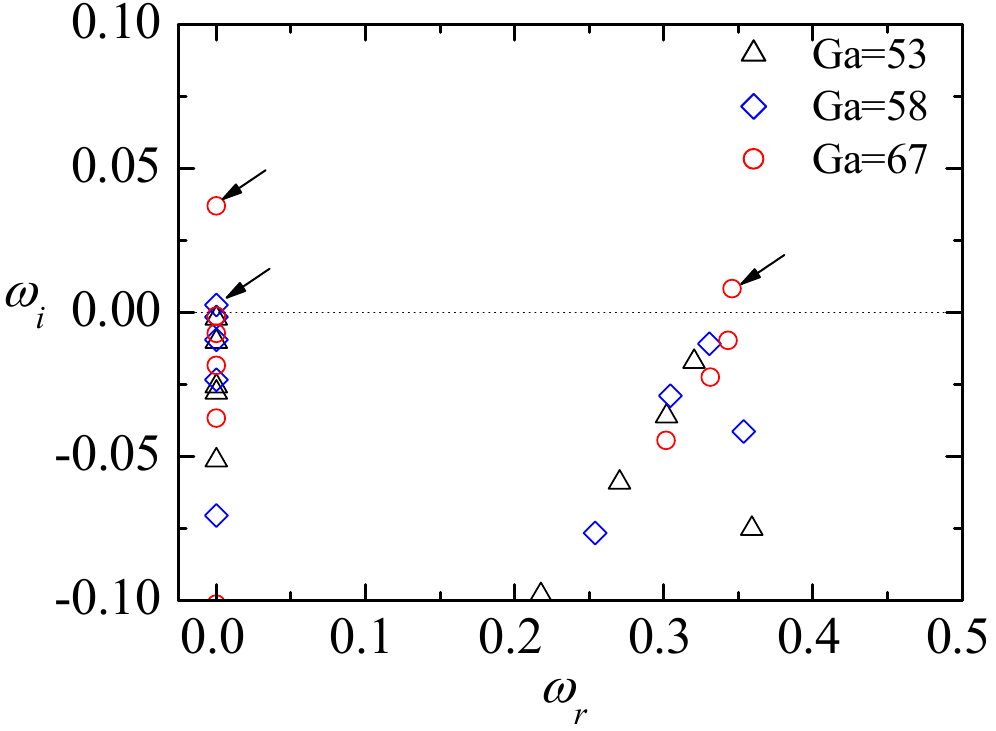}}
\end{center}
\caption{Eigenvalues for $0\leq \omega_r\leq 0.5$ and $\omega_i\geq -0.1$. The results were obtained for $c_{\infty}/c_{\text{cmc}}=10^{-3}$. The arrows indicate the unstable eigenmodes. The eigenvalues are made dimensioneless with the inertio-capillary time $t_{ic}=(\rho R^3/\gamma_c)^{1/2}$.}
\label{eigen}
\end{figure}

Figure \ref{eigen} shows the eigenvalues for the three cases considered in Sec.\ \ref{sec51}. As mentioned above, the bubble follows a straight (stable) path for $\text{Ga}=53$. A stationary ($\omega_r=0$) instability arises at $\text{Ga}=58$. A secondary oscillatory Hopf bifurcation occurs at $\text{Ga}=66$. 

The surfactant practically immobilizes the bubble surface. This explains why the bubble path suffers from the same sequence of instabilities as that of a solid sphere: a stationary instability followed by the coexistence of stationary and oscillatory instabilities. This sequence differs from that observed in clean water: an oscillatory instability followed by the coexistence of oscillatory and stationary instabilities \citep{BFM23}. Interestingly, stationary modes (oblique paths) have been reported neither in experiments nor in simulations with clean water. It has been speculated that the changes introduced in the base flow by the primary oscillatory mode prevent the stationary perturbation from growing \citep{BSFM24}.

\begin{figure}
\vspace{0.cm}
\begin{center}
\resizebox{0.45\textwidth}{!}{\includegraphics{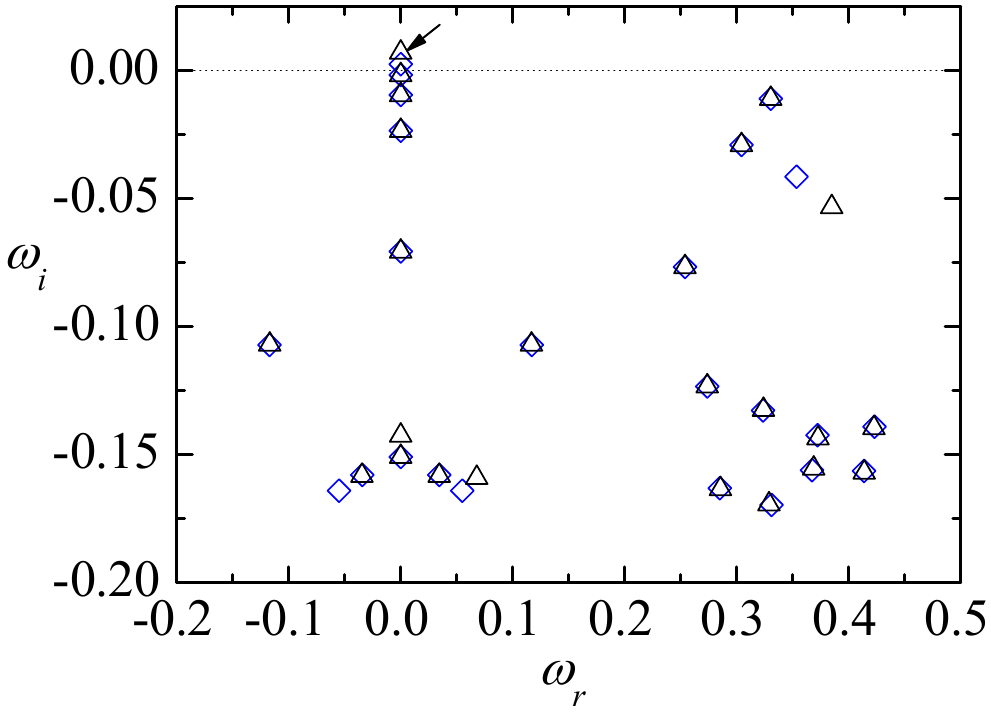}}
\end{center}
\caption{Eigenvalues for $0\leq \omega_r\leq 0.5$ and $\omega_i\geq -0.175$. The results were obtained for $c_{\infty}/c_{\text{cmc}}=10^{-3}$ and $\text{Ga}=58$. The triangles correspond to the results obtained by fixing the surfactant concentration (all the variables are perturbed except for the surfactant concentration). The eigenvalues are made dimensioneless with th inertio-capillary time $t_{ic}=(\rho R^3/\gamma_c)^{1/2}$.}
\label{solid}
\end{figure}

A fundamental difference exists between the surfactant-covered bubble and the solid sphere. The Marangoni stress responsible for the surface immobilization is perturbed due to the surfactant concentration perturbation. Conversely, the no-slip boundary condition cannot be perturbed. Figure \ref{solid} shows the effect of fixing the surfactant concentration (and, therefore, the Marangoni stress) on the eigenvalues. In this case, all the variables are perturbed except for the surfactant concentration, implying that the Marangoni stress is that of the base flow at any instant. The growth rate of the critical eigenmode increases from 0.00259 to 0.00723, implying that the flow becomes more unstable. This means that the perturbation of the surface tension over the bubble partially suppresses the critical mode growth, and points out the important role played by the perturbation of the viscous stress at the surface (that balances Marangoni stress).

\begin{figure*}
\vspace{0.cm}
\begin{center}
\resizebox{\textwidth}{!}{\includegraphics{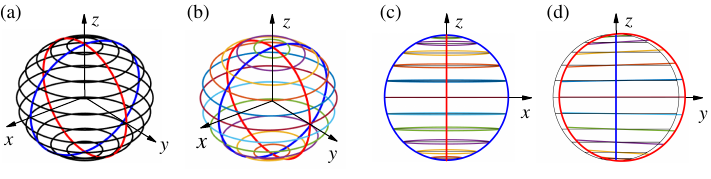}}
\end{center}
\caption{Bubble shape in the base flow indicating the analyzed parallels and meridians (a), perturbed bubble shape (b), projection of the perturbed bubble shape onto the $xz$ (c) and $yz$ (d) planes. The results were obtained for $c_{\infty}/c_{\text{cmc}}=10^{-3}$ and $\text{Ga}=58$.}
\label{ShapePer}
\end{figure*}

Figure \ref{ShapePer} shows how the bubble shape is perturbed by the critical mode of the unstable case $\text{Ga}=58$ and $c_{\infty}/c_{\text{cmc}}=10^{-3}$. Several parallels have been plotted to make the effect more visible. The deformation has been adapted for visualization. The axes have been oriented so that the shape is not perturbed in the $xz$ plane, and the maximum perturbation occurs in the $yz$ plane. For all quantities analyzed, the perturbation is symmetric with respect to the $yz$ plane. Since the bubble is axisymmetric in the base flow, the projections of the parallels onto the $xz$ and $yz$ planes are straight lines. The perturbation produces the bubble displacement to the right (the positive $y$ axis direction) (Fig.\ \ref{ShapePer}d) and the bubble deformation due to the rotation of the parallels around the $x$ axis (Fig.\ \ref{ShapePer}c-d). This rotation is anticlockwise for the equator and the parallels above it. The rotation changes to clockwise slightly below the equator. As a result, the bubble surface expands in the displacement direction and shrinks in the opposite direction. 

\begin{figure*}
\vspace{0.cm}
\begin{center}
\resizebox{0.95\textwidth}{!}{\includegraphics{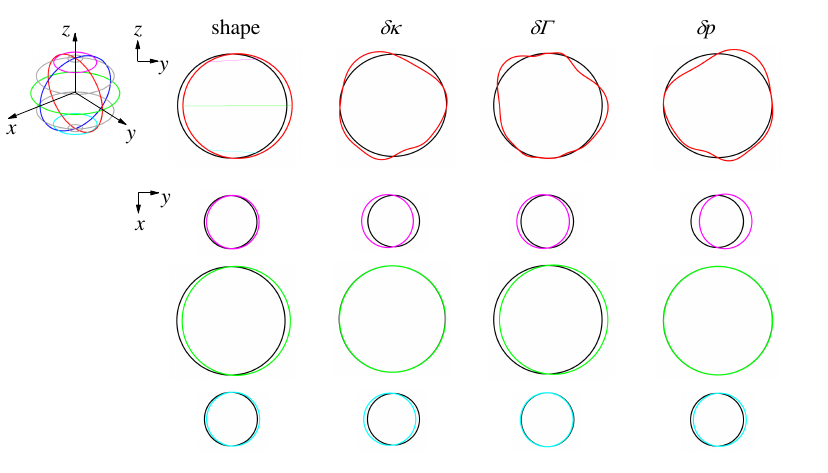}}
\end{center}
\caption{Perturbation of the bubble shape, curvature $\delta \kappa$, surfactant surface concentration $\delta\Gamma$ ($-\delta\gamma$), and hydrostatic pressure $\delta p$ in different sections of the bubble surface. The magnitudes have been adapted for visualization. The sphere in the upper-left corner indicates the meridian and the parallels considered in the figure. The results were obtained for $c_{\infty}/c_{\text{cmc}}=10^{-3}$ and $\text{Ga}=58$.}
\label{Fig_Pertub01}
\end{figure*}

As shown below, the normal viscous stress associated with the perturbation practically vanishes for $\text{Ga}=58$. This means that the hydrostatic pressure perturbation $\delta p$ practically equals $-\delta p_c$, where $p_c=\gamma\, \kappa$ is the capillary pressure. Therefore, the hydrostatic pressure is perturbed due to changes in the surface tension $\gamma$ and curvature $\kappa$, i.e., $\delta p\simeq -(\delta \gamma\kappa_0+\gamma_0\delta\kappa)$. These two contributions are shown in Fig.\ \ref{Fig_Pertub01}. The comparison between $\delta p$ and $\delta\kappa$ suggests that $\gamma_0\, \delta\kappa$ constitutes the main contribution to $\delta p$. In other words, the changes in the surface tension contribute to the hydrostatic pressure perturbation on the bubble surface to a lesser extent. The expansion of the bubble's right side (Fig.\ \ref{ShapePer}) produces a reduction in the curvature, which contributes to decreasing the capillary pressure (increasing the hydrostatic pressure) (Fig.\ \ref{Fig_Pertub01}). The opposite occurs on the bubble's left side. This effect is more significant at the bubble front where the base flow surfactant concentration is lower ($\gamma_0$ is higher).

\begin{figure*}
\vspace{0.cm}
\begin{center}
\resizebox{0.245\textwidth}{!}{\includegraphics{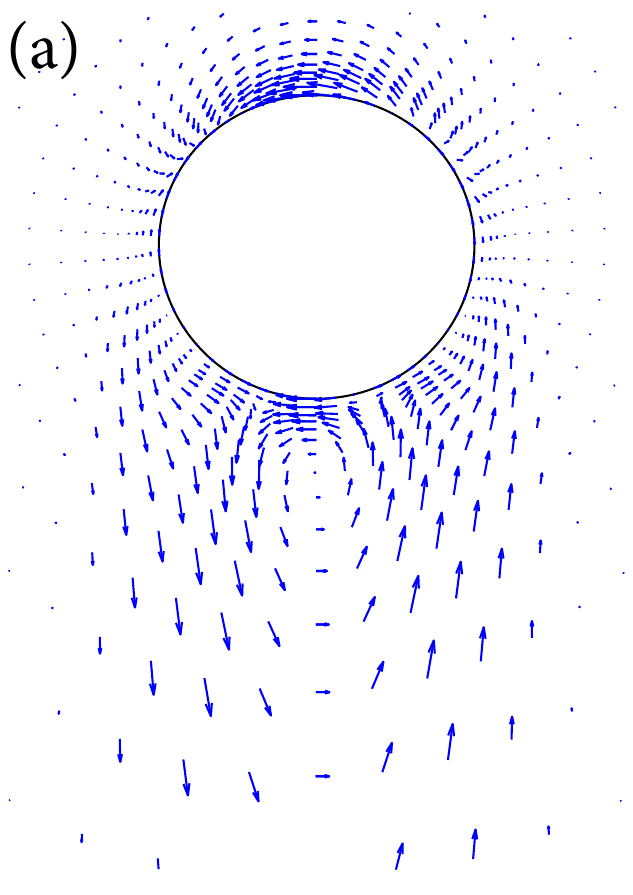}}\resizebox{0.245\textwidth}{!}{\includegraphics{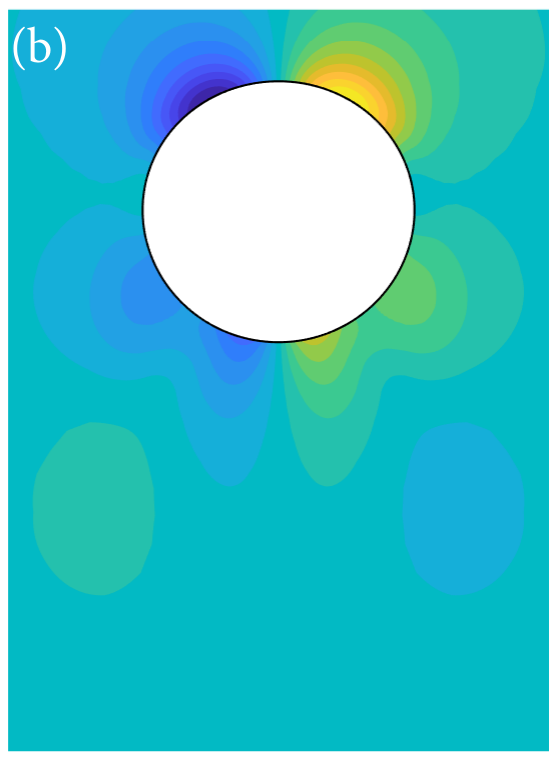}}
\resizebox{0.245\textwidth}{!}{\includegraphics{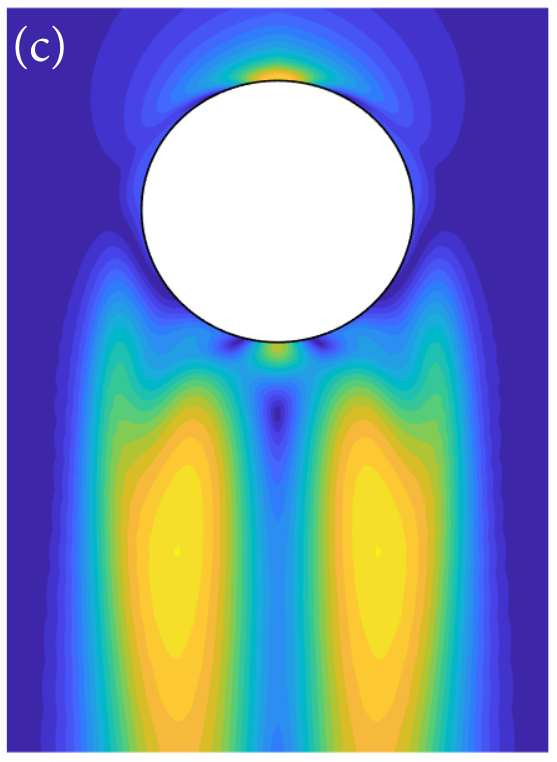}}\resizebox{0.245\textwidth}{!}{\includegraphics{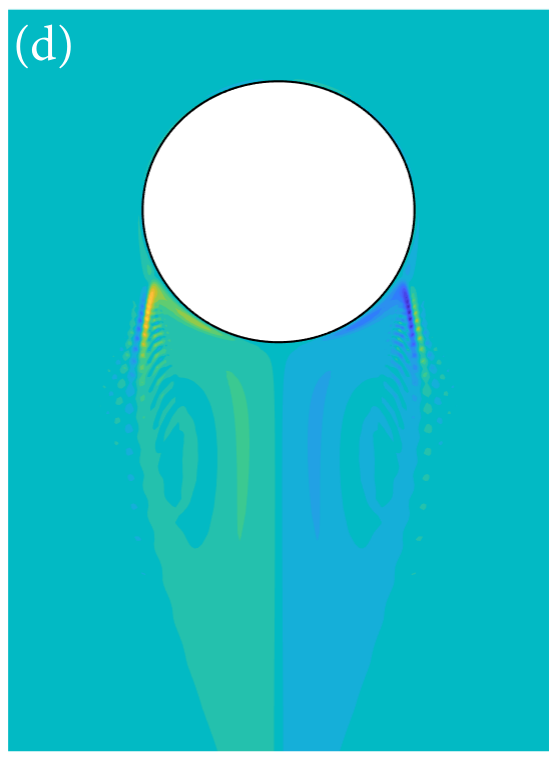}}
\end{center}
\caption{Perturbation in the $yz$ plane of (a) the velocity projection onto the $yz$ plane, (b) the hydrostatic pressure, (c) the magnitude of the velocity perturbation, and (d) the surfactant volumetric concentration. Yellow (blue) corresponds to higher (lower) values of the corresponding quantity. The results were obtained for $c_{\infty}/c_{\text{cmc}}=10^{-3}$ and $\text{Ga}=58$.}
\label{Fig_Perturb02}
\end{figure*}

Figure \ref{Fig_Perturb02}a shows the projection onto the $yz$ plane of the velocity field perturbation evaluated in that plane. The horizontal pressure gradient (Fig.\ \ref{Fig_Perturb02}b) pushes the liquid surrounding the bubble to the left at both the front and rear of the bubble. The magnitude of the velocity perturbation (Fig.\ \ref{Fig_Perturb02}c) increases in the front and rear of the bubble and reaches its maximum value in the wake. In an experiment, the velocity perturbation remains small in terms of the terminal velocity. However, it can be significant compared to the surface velocity because the interface is almost immobilized. Therefore, the flow perturbation can alter the delicate balance between the surfactant transport mechanisms described in Sec.\ \ref{sec52}. The pressure field perturbation (Fig.\ \ref{Fig_Perturb02}b) reflects the hydrostatic pressure perturbation $\delta p$ discussed above (Fig.\ \ref{Fig_Pertub01}). Figure \ref{Fig_Perturb02}d shows that the perturbation of the surfactant volumetric concentration is antisymmetric with respect to the $yz$ plane, as can be anticipated from $\delta\Gamma$ (Fig.\ \ref{Fig_Pertub01}). The perturbation of the surfactant volumetric concentration is confined within the wake.

The capillary pressure is also affected by the surface tension perturbation $\delta \gamma$ due to changes in the surfactant surface concentration. This concentration is altered by the surface expansion/contraction and the surfactant flux produced by the perturbed flow. The general view of the velocity field perturbation shows an anticlockwise rotation of the liquid around the bubble and the formation of a recirculation cell rotating in the opposite direction in the bubble rear (Fig. \ref{Fig_Perturb02}a). 

\begin{figure*}
\vspace{0.cm}
\begin{center}
\resizebox{0.65\textwidth}{!}{\includegraphics{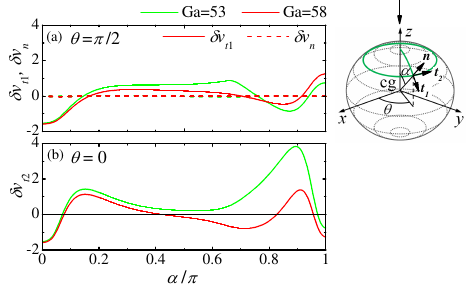}}
\end{center}
\caption{Components $\delta v_n$, $\delta v_{t1}$, and $\delta v_{t2}$ of the velocity perturbation at the bubble surface. The results were obtained for $c_{\infty}/c_{\text{cmc}}=10^{-3}$ and $\text{Ga}=53$ and 58.}
\label{Fig_velocity}
\end{figure*}

Figure \ref{Fig_velocity} shows the three components of the velocity field perturbation on the bubble surface. Here, $\delta v_n$, $\delta v_{t1}$, and $\delta v_{t2}$ are the velocity components along the normal and two tangential directions to the bubble surface, respectively. We plot the results evaluated at the meridians $\theta=0$ and $\pi/2$, as indicated in the figure. The normal component $\delta v_n$ is insignificant compared to the tangential component. The surfactant is dragged toward the left side at the bubble front ($\alpha/\pi\lesssim 0.1$) and rear ($\alpha/\pi\gtrsim 0.95$), as shown by the values of $\delta v_{t1}$ and $\delta v_{t2}$. However, the surface contraction and the incoming surfactant flux increase the concentration on the left side (Fig. \ref{Fig_Pertub01}), reducing the surface tension and decreasing the capillary pressure against the effect of the curvature increase. Furthermore, the horizontal flow (the tangential component $\delta v_{t2}$) switches direction to the right due to the significant horizontal gradient in the concentration of the surfactant. At the equator, there is almost no pressure change as the effect of the curvature and surface concentration perturbation practically compensate for each other (Fig. \ref{Fig_Pertub01}). Interestingly, the angular direction of the horizontal flow (the sign of $\delta v_{t2}$) changes several times along the bubble height (the value of $\alpha$) for $\text{Ga}=58$. 

\begin{figure*}
\vspace{0.cm}
\begin{center}
\resizebox{0.25\textwidth}{!}{\includegraphics{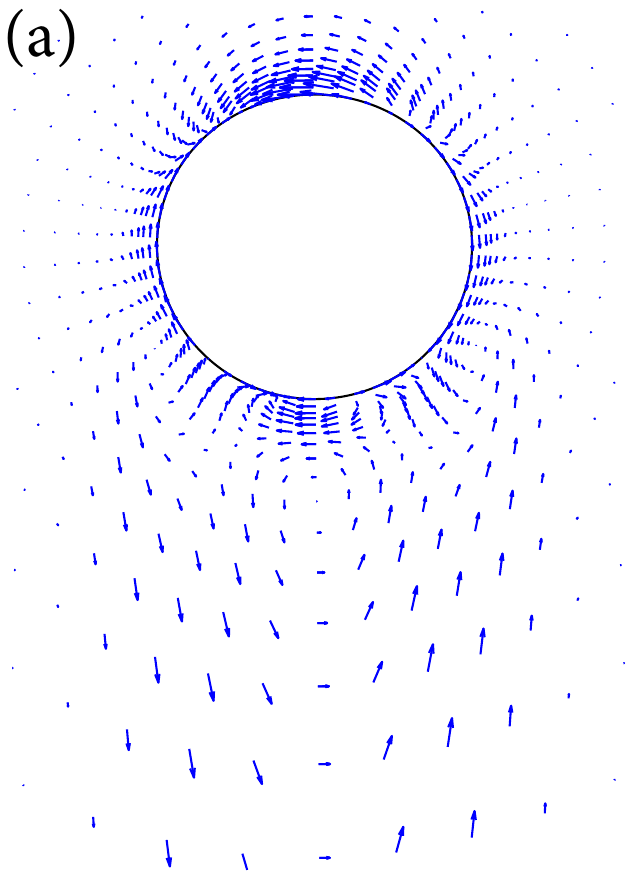}}
\resizebox{0.25\textwidth}{!}{\includegraphics{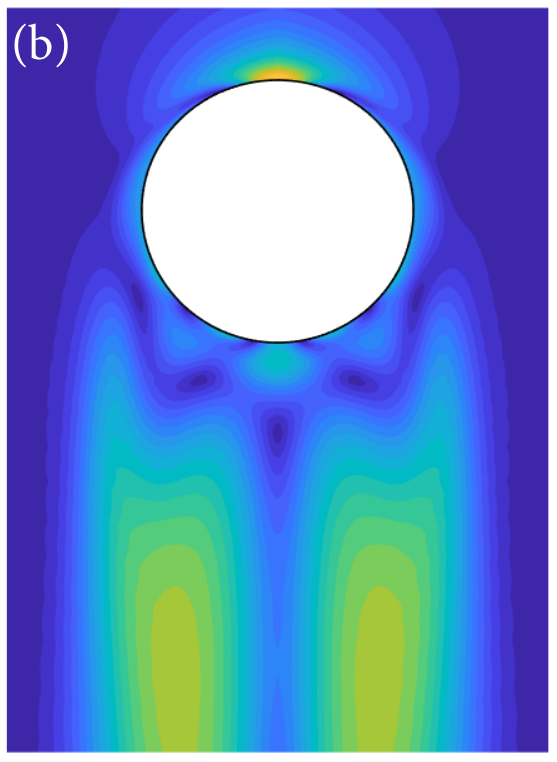}}
\end{center}
\caption{(a) Perturbation in the $yz$ plane of the velocity projection on the $yz$ plane. (b) Magnitude of the velocity field perturbation. Yellow (blue) corresponds to higher (lower) values. The results were obtained for $c_{\infty}/c_{\text{cmc}}=10^{-3}$ and $\text{Ga}=53$.}
\label{Fig_PerturbG53}
\end{figure*}

The surface forces exerted by the liquid on the bubble are altered by the bubble deformation and by the perturbation of the hydrostatic pressure and viscous stresses on the bubble surface. The equivalent system of perturbed forces is a lateral force acting in the $-y$ direction and a clockwise torque ($-x$). As shown below, this lateral force and torque have the same directions in the stable ($\text{Ga}=53$) and unstable ($\text{Ga}=58$) cases. Specifically, the lateral force pushes the bubble back towards the original position, while the torque reaction exerted on the liquid contributes to the rotating perturbed flow shown in Fig.\ \ref{Fig_Perturb02} and  \ref{Fig_PerturbG53} (note that the torque exerted on the liquid is anticlockwise). Therefore, the instability must be explained by a local effect.

\begin{figure*}
\vspace{0.cm}
\begin{center}
\resizebox{0.65\textwidth}{!}{\includegraphics{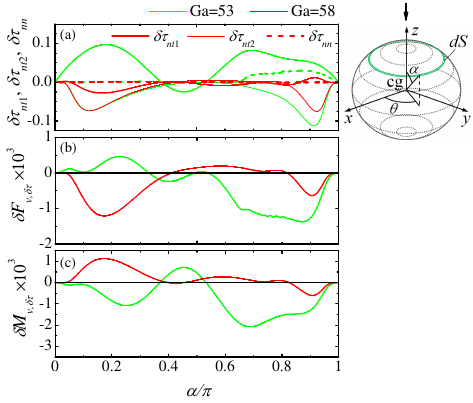}}
\end{center}
\caption{(a) Tangential $\delta \tau_{nt1}$ and normal $\delta \tau_{nn}$ elements of the perturbed viscous stress tensor evaluated at the meridian $\theta=\pi/2$, and tangential element $\delta \tau_{nt2}$ evaluated at $\theta=0$. Contributions to the viscous lateral force $\delta F_{v,\delta \tau}$ (b) and torque $\delta M_{v,\delta \tau}$ (c) resulting from the perturbed viscous stress tensor. The results were obtained for $c_{\infty}/c_{\text{cmc}}=10^{-3}$ and $\text{Ga}=53$ and 58.}
\label{Fig_ForcesAndTorques}
\end{figure*}

To analyze the source of instability, we calculated the contributions of the pressure ($\delta F_p(\alpha)$) and viscous stress ($\delta F_v(\alpha)$) perturbations to the lateral force ($\delta F(\alpha)$) exerted on the annular surface element of area $dS$ (Fig.\ \ref{Fig_ForcesAndTorques}) as
\begin{eqnarray}
\label{for}
&&\delta F=\delta F_p+\delta F_v, \quad \delta F_p=\delta F_{p,\delta p}+\delta F_{p,sh}, \nonumber\\ 
&&\delta F_v=\delta F_{v,\delta \tau}+\delta F_{v,sh},
\end{eqnarray}
where $\delta F_{p,\delta p}$ and $\delta F_{p,sh}$ stand for the contributions of the pressure and bubble shape perturbations to $\delta F_p$. Analogously, $\delta F_{v,\delta \tau}$ and $\delta F_{v,sh}$ are the contributions of the viscous stress and bubble shape perturbations to $\delta F_v$. We proceeded in a similar way to determine the $x$ component of the torque ($\delta M(\alpha)$):
\begin{eqnarray}
&&\delta M=\delta M_p+\delta M_v, \quad \delta M_p=\delta M_{p,\delta p}+\delta M_{p,sh}, \nonumber \\
&&\delta M_v=\delta M_{v,\delta \tau}+\delta M_{v,sh},
\end{eqnarray}
where the meaning of the subindexes is the same as that in Eqs.\ (\ref{for}). Figure 4 in the Supplemental Material shows all the results. Here, we summarize them. For the stable and unstable cases, $\delta F_{p,\delta p}$ is the main contribution (approximately 80\%) to the lateral force $\delta F$. In addition, $\delta M_v$ represents the major contribution (approximately 80\%) to $\delta M$. There is a noticeable reduction in $\delta M$ in the unstable case. Now, we analyze the critical role played by the contribution $\delta M_{v,\delta \tau}$ to this reduction.

Figure \ref{Fig_ForcesAndTorques}a shows the components of the viscous stress on the bubble surface. The tangential components balance the Marangoni stress resulting from the perturbation of the surfactant concentration. For both values of $\text{Ga}$, the tangential component $\delta \tau_{nt2}$ opposes the horizontal flow to the interface right side (the expanded and less loaded side). However, the component $\delta \tau_{nt1}$ switches direction and shows a smaller magnitude in the unstable case. As mentioned above, $\delta \tau_{nn}\simeq 0$ for $\text{Ga}=58$, meaning that $\delta p\simeq \delta p_c$. For $\text{Ga}=53$, there is significant normal stress in the southern hemisphere associated with the presence of secondary recirculation cells in the perturbation wake (Fig. \ref{Fig_PerturbG53}a). Figure \ref{Fig_ForcesAndTorques}b shows the lateral force $\delta F_{v,\delta \tau}$ resulting from the action of the viscous stress. For $\text{Ga}=58$, the effects of the two tangential components add up in the northern hemisphere and are counteracted in the southern one. The opposite occurs for $\text{Ga}=53$. In this case, the normal stress contributes to the lateral force in the southern hemisphere. 

Figure \ref{Fig_ForcesAndTorques}c shows the $x$ component of the torque caused by the viscous stress perturbation, $\delta M_{v,\delta \tau}$. For the stable case $\text{Ga}=53$, the integral $\Phi=\int_0^{\pi} \delta M_{v,\delta \tau} d\alpha$ is negative, which implies that the viscous stress perturbation favors the anticlockwise overall rotation of the perturbed flow around the bubble (Fig.\ \ref{Fig_PerturbG53}). In contrast, $\Phi$ is positive for the unstable case, producing the reduction of the net torque. In both cases, $\delta M_{v,\delta \tau}$ opposes the recirculation cell at the rear (see, e.g., Fig.\ \ref{Fig_PerturbG53}a), limiting the magnitude of the velocity there (see, e.g., Fig.\ \ref{Fig_PerturbG53}b). However, the torque opposing the recirculation in the rear is significantly smaller in the unstable case (Fig.\ \ref{Fig_ForcesAndTorques}c). This explains why the magnitude of the velocity field perturbation in the wake is larger than in the stable case (Fig.\ \ref{Fig_Perturb02}c). In the unstable case, more energy is transferred to the recirculation in the rear, which can produce the tilting of the main toroidal vortex of the steady flow, leading to the oblique trajectory. A similar mechanism has been described by \citet{NA93} and \citet{ERFM12} for the first bifurcation in the fixed-sphere case.

A natural question is whether the interface deformation plays a relevant role in the path stability. We do not have a definitive answer to this question. The torque $\delta M_{v,sh}$ associated with the perturbation of the interface location can be seen as the sum of two contributions: (i) that due to the bubble motion as a solid rigid and (ii) that due to the interface deformation shown in Fig.\ \ref{ShapePer}. We have verified that those two contributions are commensurate with each other, indicating that the interface deformation produces a significant effect on the torque. However, our qualitative explanation of the instability is not influenced by the interface deformation effect, which might suggest that this deformation does not significantly alter the instability mechanism.  

\section{Conclusions}

We have conducted the global stability analysis of a bubble rising in the presence of a soluble surfactant. Using a fast surfactant has allowed us to identify a parameter window in which the instability transition occurs without sharp gradients of the surfactant surface concentration, enabling an accurate calculation of the eigenmodes. In addition, the only surfactant properties entering the problem are the depletion length and the maximum packing concentration, which can be readily determined from surface tension measurements at equilibrium. The global stability analysis predictions agree with the experimental results without fitting any parameters.  

The Marangoni stress almost immobilizes the interface. The surface velocity takes tiny values as compared with the bubble terminal velocity. However, the non-zero surface velocity is crucial to understanding the surfactant behavior. Without surface convection, diffusion would render surfactant concentration uniform, suppressing the Marangoni stress that immobilizes the interface. The base flow for the subcritical and supercritical conditions is qualitatively the same. The instability mechanism is explained in terms of the effect of the critical eigenmode perturbation on the bubble path.

For the surfactant concentration considered in our analysis, the bubble path suffers from a stationary instability (oblique path) above a Galilei number threshold, as occurs in the experiments. The critical eigenmode deforms the bubble, which expands in the displacement direction and shrinks in the opposite direction. The hydrostatic pressure perturbation on the bubble surface is mainly caused by the curvature perturbation. The perturbation of the surfactant concentration is confined within the wake and is practically antisymmetric with respect to the $yz$ plane ($y$ and $z$ are the horizontal and vertical displacement directions, respectively). The perturbed viscous stress produces a stabilizing torque opposing the recirculation in the rear bubble. This torque significantly decreases above the critical Galilei number, which may explain the tilting of the main toroidal vortex of the steady flow.

Our experimental study has shown that considerable differences exist between the rising of bubbles in the presence of SDS and Surfynol. These differences must be attributed to the values of $\Lambda_d$ and the sorption rates in both cases. Although the sorption rate for Surfynol is unknown, it can be assumed to be much higher than for standard surfactants, such as SDS and Triton X-100 \citep{VSQCC24}. In fact, overshooting occurs in a much shorter time interval with Surfynol, indicating that desorption takes place much faster. Our results suggest that bubble rising may be used as a testbed to determine the desorption rate of a fast surfactant.

Here, we summarize the peculiarities of Surfynol found in our experimental analysis. In experiments with surfactants, one can identify three regimes as the concentration increases: the nearly clean bubble (high Re) and the immobilized interface (low Re) modes, and an intermediate regime between those modes (intermediate values of Re). In the case of Surfynol, the immobilized interface mode is reached at much smaller concentrations, and the transition from the clean bubble to the immobilized interface behavior occurs within a narrower interval of the concentration. Unstable realizations are found only at low Re, similar to those of the immobilized interface mode. Conversely, the path of bubbles loaded with SDS can become unstable for intermediate values of Re. In the presence of Surfynol, the stability character can exhibit a non-monotonous dependence with respect to the surfactant concentration for a fixed value of the bubble radius. Finally, instability can arise at a critical bubble radius within the immobilized interface regime, considerably simplifying the global stability analysis.

\vspace{1cm}

{\bf Declaration of Interests}. The authors report no conflict of interest.

\vspace{1cm}
{\bf Acknowledgement.} We gratefully acknowledge support from the Spanish Ministry of Science and Innovation (MCIN) (grant no. PID2022-140951OB-C21 and PID2022-140951OB-C22/ AEI/10.13039/501100011033/ FEDER, UE), Gobierno de Extremadura (grant no. GR21091). DF-M acknowledges grant PREP2022-000205 funded by MICIU/AEI /10.13039/501100011033 and ESF+. 


\begin{thebibliography}{56}%
\makeatletter
\providecommand \@ifxundefined [1]{%
 \@ifx{#1\undefined}
}%
\providecommand \@ifnum [1]{%
 \ifnum #1\expandafter \@firstoftwo
 \else \expandafter \@secondoftwo
 \fi
}%
\providecommand \@ifx [1]{%
 \ifx #1\expandafter \@firstoftwo
 \else \expandafter \@secondoftwo
 \fi
}%
\providecommand \natexlab [1]{#1}%
\providecommand \enquote  [1]{``#1''}%
\providecommand \bibnamefont  [1]{#1}%
\providecommand \bibfnamefont [1]{#1}%
\providecommand \citenamefont [1]{#1}%
\providecommand \href@noop [0]{\@secondoftwo}%
\providecommand \href [0]{\begingroup \@sanitize@url \@href}%
\providecommand \@href[1]{\@@startlink{#1}\@@href}%
\providecommand \@@href[1]{\endgroup#1\@@endlink}%
\providecommand \@sanitize@url [0]{\catcode `\\12\catcode `\$12\catcode `\&12\catcode `\#12\catcode `\^12\catcode `\_12\catcode `\%12\relax}%
\providecommand \@@startlink[1]{}%
\providecommand \@@endlink[0]{}%
\providecommand \url  [0]{\begingroup\@sanitize@url \@url }%
\providecommand \@url [1]{\endgroup\@href {#1}{\urlprefix }}%
\providecommand \urlprefix  [0]{URL }%
\providecommand \Eprint [0]{\href }%
\providecommand \doibase [0]{http://dx.doi.org/}%
\providecommand \selectlanguage [0]{\@gobble}%
\providecommand \bibinfo  [0]{\@secondoftwo}%
\providecommand \bibfield  [0]{\@secondoftwo}%
\providecommand \translation [1]{[#1]}%
\providecommand \BibitemOpen [0]{}%
\providecommand \bibitemStop [0]{}%
\providecommand \bibitemNoStop [0]{.\EOS\space}%
\providecommand \EOS [0]{\spacefactor3000\relax}%
\providecommand \BibitemShut  [1]{\csname bibitem#1\endcsname}%
\let\auto@bib@innerbib\@empty
\bibitem [{\citenamefont {Saffman}(1956)}]{S56}%
  \BibitemOpen
  \bibfield  {author} {\bibinfo {author} {\bibfnamefont {P.G.}\ \bibnamefont {Saffman}},\ }\bibfield  {title} {\enquote {\bibinfo {title} {On the rise of small air bubbles in water},}\ }\href@noop {} {\bibfield  {journal} {\bibinfo  {journal} {J. Fluid Mech.}\ }\textbf {\bibinfo {volume} {1}},\ \bibinfo {pages} {249--275} (\bibinfo {year} {1956})}\BibitemShut {NoStop}%
\bibitem [{\citenamefont {Sanada}\ \emph {et~al.}(2008)\citenamefont {Sanada}, \citenamefont {Sugihara}, \citenamefont {Shirota},\ and\ \citenamefont {Watanabe}}]{SSSW08}%
  \BibitemOpen
  \bibfield  {author} {\bibinfo {author} {\bibfnamefont {T.}~\bibnamefont {Sanada}}, \bibinfo {author} {\bibfnamefont {K.}~\bibnamefont {Sugihara}}, \bibinfo {author} {\bibfnamefont {M.}~\bibnamefont {Shirota}}, \ and\ \bibinfo {author} {\bibfnamefont {M.}~\bibnamefont {Watanabe}},\ }\bibfield  {title} {\enquote {\bibinfo {title} {Motion and drag of a single bubble in super-purified water},}\ }\href@noop {} {\bibfield  {journal} {\bibinfo  {journal} {Fluid Dynam. Res.}\ }\textbf {\bibinfo {volume} {40}},\ \bibinfo {pages} {534--545} (\bibinfo {year} {2008})}\BibitemShut {NoStop}%
\bibitem [{\citenamefont {Tripathi}\ \emph {et~al.}(2015)\citenamefont {Tripathi}, \citenamefont {Sahu},\ and\ \citenamefont {Govindarajan}}]{TSG15}%
  \BibitemOpen
  \bibfield  {author} {\bibinfo {author} {\bibfnamefont {M.~K.}\ \bibnamefont {Tripathi}}, \bibinfo {author} {\bibfnamefont {K.~C.}\ \bibnamefont {Sahu}}, \ and\ \bibinfo {author} {\bibfnamefont {R.}~\bibnamefont {Govindarajan}},\ }\bibfield  {title} {\enquote {\bibinfo {title} {Dynamics of an initially spherical bubble rising in quiescent liquid},}\ }\href@noop {} {\bibfield  {journal} {\bibinfo  {journal} {Nat Commun}\ }\textbf {\bibinfo {volume} {6}},\ \bibinfo {pages} {6268} (\bibinfo {year} {2015})}\BibitemShut {NoStop}%
\bibitem [{\citenamefont {Cano-Lozano}\ \emph {et~al.}(2016{\natexlab{a}})\citenamefont {Cano-Lozano}, \citenamefont {Tchoufag}, \citenamefont {Magnaudet},\ and\ \citenamefont {Mart\'{\i}nez-Baz\'an}}]{CTMM16}%
  \BibitemOpen
  \bibfield  {author} {\bibinfo {author} {\bibfnamefont {J.C.}\ \bibnamefont {Cano-Lozano}}, \bibinfo {author} {\bibfnamefont {J.}~\bibnamefont {Tchoufag}}, \bibinfo {author} {\bibfnamefont {J.}~\bibnamefont {Magnaudet}}, \ and\ \bibinfo {author} {\bibfnamefont {C.}~\bibnamefont {Mart\'{\i}nez-Baz\'an}},\ }\bibfield  {title} {\enquote {\bibinfo {title} {A global stability approach to wake and path instabilities of nearly oblate spheroidal rising bubbles},}\ }\href@noop {} {\bibfield  {journal} {\bibinfo  {journal} {Phys. Fluids}\ }\textbf {\bibinfo {volume} {198}},\ \bibinfo {pages} {014102} (\bibinfo {year} {2016}{\natexlab{a}})}\BibitemShut {NoStop}%
\bibitem [{\citenamefont {Cano-Lozano}\ \emph {et~al.}(2016{\natexlab{b}})\citenamefont {Cano-Lozano}, \citenamefont {Mart\'{\i}nez-Baz\'an}, \citenamefont {Magnaudet},\ and\ \citenamefont {Tchoufag}}]{CMMT16}%
  \BibitemOpen
  \bibfield  {author} {\bibinfo {author} {\bibfnamefont {J.~C.}\ \bibnamefont {Cano-Lozano}}, \bibinfo {author} {\bibfnamefont {C.}~\bibnamefont {Mart\'{\i}nez-Baz\'an}}, \bibinfo {author} {\bibfnamefont {J.}~\bibnamefont {Magnaudet}}, \ and\ \bibinfo {author} {\bibfnamefont {J.}~\bibnamefont {Tchoufag}},\ }\bibfield  {title} {\enquote {\bibinfo {title} {Paths and wakes of deformable nearly spheroidal rising bubbles close to the transition to path instability},}\ }\href@noop {} {\bibfield  {journal} {\bibinfo  {journal} {Phys. Rev. Fluids}\ }\textbf {\bibinfo {volume} {1}},\ \bibinfo {pages} {053604} (\bibinfo {year} {2016}{\natexlab{b}})}\BibitemShut {NoStop}%
\bibitem [{\citenamefont {Kure}\ \emph {et~al.}(2021)\citenamefont {Kure}, \citenamefont {Jakobsen}, \citenamefont {{La Forgia}},\ and\ \citenamefont {Solsvik}}]{KJLS21}%
  \BibitemOpen
  \bibfield  {author} {\bibinfo {author} {\bibfnamefont {I.~K.}\ \bibnamefont {Kure}}, \bibinfo {author} {\bibfnamefont {H.~A.}\ \bibnamefont {Jakobsen}}, \bibinfo {author} {\bibfnamefont {N.}~\bibnamefont {{La Forgia}}}, \ and\ \bibinfo {author} {\bibfnamefont {J.}~\bibnamefont {Solsvik}},\ }\bibfield  {title} {\enquote {\bibinfo {title} {Experimental investigation of single bubbles rising in stagnant liquid: {Statistical} analysis and image processing},}\ }\href@noop {} {\bibfield  {journal} {\bibinfo  {journal} {Phys. Fluids}\ }\textbf {\bibinfo {volume} {33}},\ \bibinfo {pages} {103611} (\bibinfo {year} {2021})}\BibitemShut {NoStop}%
\bibitem [{\citenamefont {Bonnefis}\ \emph {et~al.}(2023)\citenamefont {Bonnefis}, \citenamefont {Fabre},\ and\ \citenamefont {Magnaudet}}]{BFM23}%
  \BibitemOpen
  \bibfield  {author} {\bibinfo {author} {\bibfnamefont {P.}~\bibnamefont {Bonnefis}}, \bibinfo {author} {\bibfnamefont {D.}~\bibnamefont {Fabre}}, \ and\ \bibinfo {author} {\bibfnamefont {J.}~\bibnamefont {Magnaudet}},\ }\bibfield  {title} {\enquote {\bibinfo {title} {When, how, and why the path of an air bubble rising in pure water becomes unstable},}\ }\href@noop {} {\bibfield  {journal} {\bibinfo  {journal} {Proc. Natl. Acad. Sci.}\ }\textbf {\bibinfo {volume} {120}},\ \bibinfo {pages} {e2300897120} (\bibinfo {year} {2023})}\BibitemShut {NoStop}%
\bibitem [{\citenamefont {Bonnefis}(2019)}]{B19b}%
  \BibitemOpen
  \bibfield  {author} {\bibinfo {author} {\bibfnamefont {P.}~\bibnamefont {Bonnefis}},\ }\bibfield  {title} {\enquote {\bibinfo {title} {Étude des instabilités de sillage, de forme et de trajectoire de bulles par une étude stabilité linéaire globale},}\ }\href@noop {} {\bibfield  {journal} {\bibinfo  {journal} {Ph. D. Thesis}\ } (\bibinfo {year} {2019})}\BibitemShut {NoStop}%
\bibitem [{\citenamefont {Herrada}\ and\ \citenamefont {Eggers}(2023)}]{HE23}%
  \BibitemOpen
  \bibfield  {author} {\bibinfo {author} {\bibfnamefont {M.A.}\ \bibnamefont {Herrada}}\ and\ \bibinfo {author} {\bibfnamefont {J.G.}\ \bibnamefont {Eggers}},\ }\bibfield  {title} {\enquote {\bibinfo {title} {Leonardo’s paradox resolved: path instability of an air bubble rising in water},}\ }\href@noop {} {\bibfield  {journal} {\bibinfo  {journal} {Proc. Natl. Acad. Sci.}\ }\textbf {\bibinfo {volume} {120}},\ \bibinfo {pages} {e2216830120} (\bibinfo {year} {2023})}\BibitemShut {NoStop}%
\bibitem [{\citenamefont {Natarajan}\ and\ \citenamefont {Acrivos}(1993)}]{NA93}%
  \BibitemOpen
  \bibfield  {author} {\bibinfo {author} {\bibfnamefont {R.}~\bibnamefont {Natarajan}}\ and\ \bibinfo {author} {\bibfnamefont {A.}~\bibnamefont {Acrivos}},\ }\bibfield  {title} {\enquote {\bibinfo {title} {The instability of the steady flow past spheres and disks},}\ }\href@noop {} {\bibfield  {journal} {\bibinfo  {journal} {J Fluid Mech.}\ }\textbf {\bibinfo {volume} {254}},\ \bibinfo {pages} {323--344} (\bibinfo {year} {1993})}\BibitemShut {NoStop}%
\bibitem [{\citenamefont {Yang}\ and\ \citenamefont {Prosperetti}(2007)}]{YP07}%
  \BibitemOpen
  \bibfield  {author} {\bibinfo {author} {\bibfnamefont {B.}~\bibnamefont {Yang}}\ and\ \bibinfo {author} {\bibfnamefont {A.}~\bibnamefont {Prosperetti}},\ }\bibfield  {title} {\enquote {\bibinfo {title} {Linear stability of the flow past a spheroidal bubble},}\ }\href@noop {} {\bibfield  {journal} {\bibinfo  {journal} {J. Fluid Mech.}\ }\textbf {\bibinfo {volume} {582}},\ \bibinfo {pages} {53--78} (\bibinfo {year} {2007})}\BibitemShut {NoStop}%
\bibitem [{\citenamefont {Ern}\ \emph {et~al.}(2012)\citenamefont {Ern}, \citenamefont {Risso}, \citenamefont {Fabre},\ and\ \citenamefont {Magnaudet}}]{ERFM12}%
  \BibitemOpen
  \bibfield  {author} {\bibinfo {author} {\bibfnamefont {P.}~\bibnamefont {Ern}}, \bibinfo {author} {\bibfnamefont {F.}~\bibnamefont {Risso}}, \bibinfo {author} {\bibfnamefont {D.}~\bibnamefont {Fabre}}, \ and\ \bibinfo {author} {\bibfnamefont {J.}~\bibnamefont {Magnaudet}},\ }\bibfield  {title} {\enquote {\bibinfo {title} {Wake-induced oscillatory paths of bodies freely rising or falling in fluids},}\ }\href@noop {} {\bibfield  {journal} {\bibinfo  {journal} {Annu. Rev. Fluid Mech.}\ }\textbf {\bibinfo {volume} {44}},\ \bibinfo {pages} {97--121} (\bibinfo {year} {2012})}\BibitemShut {NoStop}%
\bibitem [{\citenamefont {Tchoufag}\ \emph {et~al.}(2013)\citenamefont {Tchoufag}, \citenamefont {Magnaudet},\ and\ \citenamefont {Fabre}}]{TMF13}%
  \BibitemOpen
  \bibfield  {author} {\bibinfo {author} {\bibfnamefont {J.}~\bibnamefont {Tchoufag}}, \bibinfo {author} {\bibfnamefont {J.}~\bibnamefont {Magnaudet}}, \ and\ \bibinfo {author} {\bibfnamefont {D.}~\bibnamefont {Fabre}},\ }\bibfield  {title} {\enquote {\bibinfo {title} {Linear stability and sensitivity of the flow past a fixed oblate spheroidal bubble},}\ }\href@noop {} {\bibfield  {journal} {\bibinfo  {journal} {Phys. Fluids}\ }\textbf {\bibinfo {volume} {25}},\ \bibinfo {pages} {054108} (\bibinfo {year} {2013})}\BibitemShut {NoStop}%
\bibitem [{\citenamefont {Cano-Lozano}\ \emph {et~al.}(2013)\citenamefont {Cano-Lozano}, \citenamefont {Bohorquez},\ and\ \citenamefont {Mart\'{\i}nez-Baz\'an}}]{CBM13}%
  \BibitemOpen
  \bibfield  {author} {\bibinfo {author} {\bibfnamefont {J.~C.}\ \bibnamefont {Cano-Lozano}}, \bibinfo {author} {\bibfnamefont {P.}~\bibnamefont {Bohorquez}}, \ and\ \bibinfo {author} {\bibfnamefont {C.}~\bibnamefont {Mart\'{\i}nez-Baz\'an}},\ }\bibfield  {title} {\enquote {\bibinfo {title} {Wake instability of a fixed axisymmetric bubble of realistic shape},}\ }\href@noop {} {\bibfield  {journal} {\bibinfo  {journal} {Int. J. Multiphase Flow}\ }\textbf {\bibinfo {volume} {51}},\ \bibinfo {pages} {11--21} (\bibinfo {year} {2013})}\BibitemShut {NoStop}%
\bibitem [{\citenamefont {Tchoufag}\ \emph {et~al.}(2014)\citenamefont {Tchoufag}, \citenamefont {Magnaudet},\ and\ \citenamefont {Fabre}}]{TMF14}%
  \BibitemOpen
  \bibfield  {author} {\bibinfo {author} {\bibfnamefont {J.}~\bibnamefont {Tchoufag}}, \bibinfo {author} {\bibfnamefont {J.}~\bibnamefont {Magnaudet}}, \ and\ \bibinfo {author} {\bibfnamefont {D.}~\bibnamefont {Fabre}},\ }\bibfield  {title} {\enquote {\bibinfo {title} {Linear instability of the path of a freely rising spheroidal bubble},}\ }\href@noop {} {\bibfield  {journal} {\bibinfo  {journal} {J. Fluid Mech.}\ }\textbf {\bibinfo {volume} {751}},\ \bibinfo {pages} {R4} (\bibinfo {year} {2014})}\BibitemShut {NoStop}%
\bibitem [{\citenamefont {Bonneﬁs}\ \emph {et~al.}(2024)\citenamefont {Bonneﬁs}, \citenamefont {Sierra-Ausin}, \citenamefont {Fabre},\ and\ \citenamefont {Magnaudet}}]{BSFM24}%
  \BibitemOpen
  \bibfield  {author} {\bibinfo {author} {\bibfnamefont {P.}~\bibnamefont {Bonneﬁs}}, \bibinfo {author} {\bibfnamefont {J.}~\bibnamefont {Sierra-Ausin}}, \bibinfo {author} {\bibfnamefont {D.}~\bibnamefont {Fabre}}, \ and\ \bibinfo {author} {\bibfnamefont {J.}~\bibnamefont {Magnaudet}},\ }\bibfield  {title} {\enquote {\bibinfo {title} {Path instability of deformable bubbles rising in newtonian liquids: a linear study},}\ }\href@noop {} {\bibfield  {journal} {\bibinfo  {journal} {J. Fluid Mech.}\ }\textbf {\bibinfo {volume} {980}},\ \bibinfo {pages} {A19} (\bibinfo {year} {2024})}\BibitemShut {NoStop}%
\bibitem [{\citenamefont {Zhou}\ and\ \citenamefont {Dusek}(2017)}]{ZD17}%
  \BibitemOpen
  \bibfield  {author} {\bibinfo {author} {\bibfnamefont {W.}~\bibnamefont {Zhou}}\ and\ \bibinfo {author} {\bibfnamefont {J.}~\bibnamefont {Dusek}},\ }\bibfield  {title} {\enquote {\bibinfo {title} {Marginal stability curve of a deformable bubble},}\ }\href@noop {} {\bibfield  {journal} {\bibinfo  {journal} {Int. J. Multiphase Flow}\ }\textbf {\bibinfo {volume} {89}},\ \bibinfo {pages} {218--227} (\bibinfo {year} {2017})}\BibitemShut {NoStop}%
\bibitem [{\citenamefont {Duineveld}(1995)}]{D95}%
  \BibitemOpen
  \bibfield  {author} {\bibinfo {author} {\bibfnamefont {P.C.}\ \bibnamefont {Duineveld}},\ }\bibfield  {title} {\enquote {\bibinfo {title} {The rise velocity and shape of bubbles in pure water at high reynolds number},}\ }\href@noop {} {\bibfield  {journal} {\bibinfo  {journal} {J. Fluid Mech.}\ }\textbf {\bibinfo {volume} {292}},\ \bibinfo {pages} {325–332} (\bibinfo {year} {1995})}\BibitemShut {NoStop}%
\bibitem [{\citenamefont {Yamamoto}\ and\ \citenamefont {Ishii}(1987)}]{YI87}%
  \BibitemOpen
  \bibfield  {author} {\bibinfo {author} {\bibfnamefont {T.}~\bibnamefont {Yamamoto}}\ and\ \bibinfo {author} {\bibfnamefont {T.}~\bibnamefont {Ishii}},\ }\bibfield  {title} {\enquote {\bibinfo {title} {Effect of surface active materials on the drag coefficients and shapes of single large gas bubbles},}\ }\href@noop {} {\bibfield  {journal} {\bibinfo  {journal} {Chem. Eng., Sci.}\ }\textbf {\bibinfo {volume} {42}},\ \bibinfo {pages} {1297--1303} (\bibinfo {year} {1987})}\BibitemShut {NoStop}%
\bibitem [{\citenamefont {Rodrigue}\ \emph {et~al.}(1996)\citenamefont {Rodrigue}, \citenamefont {De-Kee},\ and\ \citenamefont {Chan-Man-Fong}}]{RKF96}%
  \BibitemOpen
  \bibfield  {author} {\bibinfo {author} {\bibfnamefont {D.}~\bibnamefont {Rodrigue}}, \bibinfo {author} {\bibfnamefont {D.}~\bibnamefont {De-Kee}}, \ and\ \bibinfo {author} {\bibfnamefont {C.F.}\ \bibnamefont {Chan-Man-Fong}},\ }\bibfield  {title} {\enquote {\bibinfo {title} {An experimental study of the effect of surfactants on the free rise velocity of gas bubbles},}\ }\href@noop {} {\bibfield  {journal} {\bibinfo  {journal} {J. Non-Newtonian Fluid Mech.}\ }\textbf {\bibinfo {volume} {66}},\ \bibinfo {pages} {213--232} (\bibinfo {year} {1996})}\BibitemShut {NoStop}%
\bibitem [{\citenamefont {Fdhila}\ and\ \citenamefont {Duineveld}(1996)}]{FD96}%
  \BibitemOpen
  \bibfield  {author} {\bibinfo {author} {\bibfnamefont {R.~Bel}\ \bibnamefont {Fdhila}}\ and\ \bibinfo {author} {\bibfnamefont {P.~C.}\ \bibnamefont {Duineveld}},\ }\bibfield  {title} {\enquote {\bibinfo {title} {The effect of surfactant on the rise of a spherical bubble at high {Reynolds} and {Peclet} numbers},}\ }\href@noop {} {\bibfield  {journal} {\bibinfo  {journal} {Phys. Fluids}\ }\textbf {\bibinfo {volume} {8}},\ \bibinfo {pages} {310--321} (\bibinfo {year} {1996})}\BibitemShut {NoStop}%
\bibitem [{\citenamefont {Zhang}\ and\ \citenamefont {Finch}(2001)}]{ZF01}%
  \BibitemOpen
  \bibfield  {author} {\bibinfo {author} {\bibfnamefont {Y.}~\bibnamefont {Zhang}}\ and\ \bibinfo {author} {\bibfnamefont {J.~A.}\ \bibnamefont {Finch}},\ }\bibfield  {title} {\enquote {\bibinfo {title} {A note on single bubble motion in surfactant solutions},}\ }\href@noop {} {\bibfield  {journal} {\bibinfo  {journal} {J. Fluid Mech.}\ }\textbf {\bibinfo {volume} {429}},\ \bibinfo {pages} {63--66} (\bibinfo {year} {2001})}\BibitemShut {NoStop}%
\bibitem [{\citenamefont {Tzounakos}\ \emph {et~al.}(2004)\citenamefont {Tzounakos}, \citenamefont {Karamanev}, \citenamefont {Margaritis},\ and\ \citenamefont {Bergougnou}}]{TKMB04}%
  \BibitemOpen
  \bibfield  {author} {\bibinfo {author} {\bibfnamefont {A.}~\bibnamefont {Tzounakos}}, \bibinfo {author} {\bibfnamefont {D.~G.}\ \bibnamefont {Karamanev}}, \bibinfo {author} {\bibfnamefont {A.}~\bibnamefont {Margaritis}}, \ and\ \bibinfo {author} {\bibfnamefont {M.~A.}\ \bibnamefont {Bergougnou}},\ }\bibfield  {title} {\enquote {\bibinfo {title} {Effect of the surfactant concentration on the rise of gas bubbles inpower-law non-newtonian liquids},}\ }\href@noop {} {\bibfield  {journal} {\bibinfo  {journal} {Ind. Eng. Chem. Res.}\ }\textbf {\bibinfo {volume} {43}},\ \bibinfo {pages} {5790--5795} (\bibinfo {year} {2004})}\BibitemShut {NoStop}%
\bibitem [{\citenamefont {Takemura}(2005)}]{T05}%
  \BibitemOpen
  \bibfield  {author} {\bibinfo {author} {\bibfnamefont {F.}~\bibnamefont {Takemura}},\ }\bibfield  {title} {\enquote {\bibinfo {title} {Adsorption of surfactants onto the surface of a spherical rising bubble and its effect on the terminal velocity of the bubble},}\ }\href@noop {} {\bibfield  {journal} {\bibinfo  {journal} {Phys. Fluids}\ }\textbf {\bibinfo {volume} {17}},\ \bibinfo {pages} {048104} (\bibinfo {year} {2005})}\BibitemShut {NoStop}%
\bibitem [{\citenamefont {Alves}\ \emph {et~al.}(2005)\citenamefont {Alves}, \citenamefont {Orvalho},\ and\ \citenamefont {Vasconcelos}}]{AOV05}%
  \BibitemOpen
  \bibfield  {author} {\bibinfo {author} {\bibfnamefont {S.~S.}\ \bibnamefont {Alves}}, \bibinfo {author} {\bibfnamefont {S.~P.}\ \bibnamefont {Orvalho}}, \ and\ \bibinfo {author} {\bibfnamefont {J.~M.~T.}\ \bibnamefont {Vasconcelos}},\ }\bibfield  {title} {\enquote {\bibinfo {title} {Effect of bubble contamination on rise velocity and mass transfer},}\ }\href@noop {} {\bibfield  {journal} {\bibinfo  {journal} {Chem. Eng. Sci.}\ }\textbf {\bibinfo {volume} {60}},\ \bibinfo {pages} {1--9} (\bibinfo {year} {2005})}\BibitemShut {NoStop}%
\bibitem [{\citenamefont {Kulkarni}\ and\ \citenamefont {Joshi}(2005)}]{KJ05}%
  \BibitemOpen
  \bibfield  {author} {\bibinfo {author} {\bibfnamefont {A.~A.}\ \bibnamefont {Kulkarni}}\ and\ \bibinfo {author} {\bibfnamefont {J.~B.}\ \bibnamefont {Joshi}},\ }\bibfield  {title} {\enquote {\bibinfo {title} {Bubble formation and bubble rise velocity in gas-liquid systems: A review},}\ }\href@noop {} {\bibfield  {journal} {\bibinfo  {journal} {Ind. Eng. Chem. Res.}\ }\textbf {\bibinfo {volume} {44}},\ \bibinfo {pages} {5873--5931} (\bibinfo {year} {2005})}\BibitemShut {NoStop}%
\bibitem [{\citenamefont {Takagi}\ and\ \citenamefont {Matsumoto}(2011)}]{TM11}%
  \BibitemOpen
  \bibfield  {author} {\bibinfo {author} {\bibfnamefont {S.}~\bibnamefont {Takagi}}\ and\ \bibinfo {author} {\bibfnamefont {Y.}~\bibnamefont {Matsumoto}},\ }\bibfield  {title} {\enquote {\bibinfo {title} {Surfactant effects on bubble motion and bubbly flows},}\ }\href@noop {} {\bibfield  {journal} {\bibinfo  {journal} {Annu. Rev. Fluid Mech.}\ }\textbf {\bibinfo {volume} {43}},\ \bibinfo {pages} {615--636} (\bibinfo {year} {2011})}\BibitemShut {NoStop}%
\bibitem [{\citenamefont {Luo}\ \emph {et~al.}(2022)\citenamefont {Luo}, \citenamefont {Wang}, \citenamefont {Zhang}, \citenamefont {Guo}, \citenamefont {Zheng}, \citenamefont {Xiang}, \citenamefont {Liu},\ and\ \citenamefont {Liu}}]{LWZGZXLL22}%
  \BibitemOpen
  \bibfield  {author} {\bibinfo {author} {\bibfnamefont {Y.}~\bibnamefont {Luo}}, \bibinfo {author} {\bibfnamefont {Z.}~\bibnamefont {Wang}}, \bibinfo {author} {\bibfnamefont {B.}~\bibnamefont {Zhang}}, \bibinfo {author} {\bibfnamefont {K.}~\bibnamefont {Guo}}, \bibinfo {author} {\bibfnamefont {L.}~\bibnamefont {Zheng}}, \bibinfo {author} {\bibfnamefont {W.}~\bibnamefont {Xiang}}, \bibinfo {author} {\bibfnamefont {H.}~\bibnamefont {Liu}}, \ and\ \bibinfo {author} {\bibfnamefont {C.}~\bibnamefont {Liu}},\ }\bibfield  {title} {\enquote {\bibinfo {title} {Experimental study of the effect of the surfactant on the single bubble rising in stagnant surfactant solutions and a mathematical model for the bubble motion},}\ }\href@noop {} {\bibfield  {journal} {\bibinfo  {journal} {Ind. Eng. Chem. Res.}\ }\textbf {\bibinfo {volume} {61}},\ \bibinfo {pages} {9514--9527} (\bibinfo {year} {2022})}\BibitemShut {NoStop}%
\bibitem [{\citenamefont {Pang}\ \emph {et~al.}(2023)\citenamefont {Pang}, \citenamefont {Jia},\ and\ \citenamefont {Fei}}]{PJF23}%
  \BibitemOpen
  \bibfield  {author} {\bibinfo {author} {\bibfnamefont {M.}~\bibnamefont {Pang}}, \bibinfo {author} {\bibfnamefont {M.}~\bibnamefont {Jia}}, \ and\ \bibinfo {author} {\bibfnamefont {Y.}~\bibnamefont {Fei}},\ }\bibfield  {title} {\enquote {\bibinfo {title} {Experimental study on effect of surfactant and solution property on bubble rising motion},}\ }\href@noop {} {\bibfield  {journal} {\bibinfo  {journal} {J. Mol. Liq.}\ }\textbf {\bibinfo {volume} {375}},\ \bibinfo {pages} {121390} (\bibinfo {year} {2023})}\BibitemShut {NoStop}%
\bibitem [{\citenamefont {Rubio}\ \emph {et~al.}(2024)\citenamefont {Rubio}, \citenamefont {Vega}, \citenamefont {Cabezas}, \citenamefont {Montanero}, \citenamefont {López-Herrera},\ and\ \citenamefont {Herrada}}]{RVCMLH24}%
  \BibitemOpen
  \bibfield  {author} {\bibinfo {author} {\bibfnamefont {A.}~\bibnamefont {Rubio}}, \bibinfo {author} {\bibfnamefont {E.J.}\ \bibnamefont {Vega}}, \bibinfo {author} {\bibfnamefont {M.G.}\ \bibnamefont {Cabezas}}, \bibinfo {author} {\bibfnamefont {J.~M.}\ \bibnamefont {Montanero}}, \bibinfo {author} {\bibfnamefont {J.~M.}\ \bibnamefont {López-Herrera}}, \ and\ \bibinfo {author} {\bibfnamefont {M.~A.}\ \bibnamefont {Herrada}},\ }\bibfield  {title} {\enquote {\bibinfo {title} {Bubble rising in the presence of a surfactant at very low concentrations},}\ }\href@noop {} {\bibfield  {journal} {\bibinfo  {journal} {Physics of Fluids}\ }\textbf {\bibinfo {volume} {36}},\ \bibinfo {pages} {062112} (\bibinfo {year} {2024})}\BibitemShut {NoStop}%
\bibitem [{\citenamefont {Fern\'andez-Mart\'{\i}nez}\ \emph {et~al.}(2025)\citenamefont {Fern\'andez-Mart\'{\i}nez}, \citenamefont {Cabezas}, \citenamefont {L\'opez-Herrera}, \citenamefont {Herrada},\ and\ \citenamefont {Montanero}}]{FCLHM25}%
  \BibitemOpen
  \bibfield  {author} {\bibinfo {author} {\bibfnamefont {D.}~\bibnamefont {Fern\'andez-Mart\'{\i}nez}}, \bibinfo {author} {\bibfnamefont {M.~G.}\ \bibnamefont {Cabezas}}, \bibinfo {author} {\bibfnamefont {J.~M.}\ \bibnamefont {L\'opez-Herrera}}, \bibinfo {author} {\bibfnamefont {M.~A.}\ \bibnamefont {Herrada}}, \ and\ \bibinfo {author} {\bibfnamefont {J.~M.}\ \bibnamefont {Montanero}},\ }\bibfield  {title} {\enquote {\bibinfo {title} {Transient bubble rising in the presence of a surfactant at very low concentrations},}\ }\href@noop {} {\bibfield  {journal} {\bibinfo  {journal} {Int. J. Multiphase Flow}\ }\textbf {\bibinfo {volume} {188}},\ \bibinfo {pages} {105205} (\bibinfo {year} {2025})}\BibitemShut {NoStop}%
\bibitem [{\citenamefont {Dukhin}\ \emph {et~al.}(1998)\citenamefont {Dukhin}, \citenamefont {Miller},\ and\ \citenamefont {Loglio}}]{DML98}%
  \BibitemOpen
  \bibfield  {author} {\bibinfo {author} {\bibfnamefont {S.~S.}\ \bibnamefont {Dukhin}}, \bibinfo {author} {\bibfnamefont {R.}~\bibnamefont {Miller}}, \ and\ \bibinfo {author} {\bibfnamefont {G.}~\bibnamefont {Loglio}},\ }\enquote {\bibinfo {title} {Physico-chemical hydrodynamics of rising bubble},}\ \ (\bibinfo  {publisher} {Elsevier},\ \bibinfo {year} {1998})\ pp.\ \bibinfo {pages} {367--432}\BibitemShut {NoStop}%
\bibitem [{\citenamefont {Dukhin}\ \emph {et~al.}(2015)\citenamefont {Dukhin}, \citenamefont {Kovalchuk}, \citenamefont {Gochev}, \citenamefont {Lotfi}, \citenamefont {Krzan}, \citenamefont {Malysa},\ and\ \citenamefont {Miller}}]{DKGLKMM15}%
  \BibitemOpen
  \bibfield  {author} {\bibinfo {author} {\bibfnamefont {S.~S.}\ \bibnamefont {Dukhin}}, \bibinfo {author} {\bibfnamefont {V.~I.}\ \bibnamefont {Kovalchuk}}, \bibinfo {author} {\bibfnamefont {G.~G.}\ \bibnamefont {Gochev}}, \bibinfo {author} {\bibfnamefont {M.}~\bibnamefont {Lotfi}}, \bibinfo {author} {\bibfnamefont {M.}~\bibnamefont {Krzan}}, \bibinfo {author} {\bibfnamefont {K.}~\bibnamefont {Malysa}}, \ and\ \bibinfo {author} {\bibfnamefont {R.}~\bibnamefont {Miller}},\ }\bibfield  {title} {\enquote {\bibinfo {title} {Dynamics of rear stagnant cap formation at the surface of spherical bubbles rising in surfactant solutions at large {Reynolds} numbers under conditions of small {Marangoni} number and slow sorption kinetics},}\ }\href@noop {} {\bibfield  {journal} {\bibinfo  {journal} {Adv. Colloid Interface Sci.}\ }\textbf {\bibinfo {volume} {222}},\ \bibinfo {pages} {260--274} (\bibinfo {year} {2015})}\BibitemShut {NoStop}%
\bibitem [{\citenamefont {Ulaganathan}\ \emph {et~al.}(2016)\citenamefont {Ulaganathan}, \citenamefont {abd C.~Gehin-Delvalc}, \citenamefont {Leserc}, \citenamefont {Gunes},\ and\ \citenamefont {Miller}}]{UGGLGM16}%
  \BibitemOpen
  \bibfield  {author} {\bibinfo {author} {\bibfnamefont {V.}~\bibnamefont {Ulaganathan}}, \bibinfo {author} {\bibfnamefont {G.~Gochev}\ \bibnamefont {abd C.~Gehin-Delvalc}}, \bibinfo {author} {\bibfnamefont {M.E.}\ \bibnamefont {Leserc}}, \bibinfo {author} {\bibfnamefont {D.~Z.}\ \bibnamefont {Gunes}}, \ and\ \bibinfo {author} {\bibfnamefont {R.}~\bibnamefont {Miller}},\ }\bibfield  {title} {\enquote {\bibinfo {title} {Effect of ph and electrolyte concentration on rising air bubbles in $\beta$-lactoglobulin solutions},}\ }\href@noop {} {\bibfield  {journal} {\bibinfo  {journal} {Coll. Surf. A}\ }\textbf {\bibinfo {volume} {505}},\ \bibinfo {pages} {165.170} (\bibinfo {year} {2016})}\BibitemShut {NoStop}%
\bibitem [{\citenamefont {Zawala}\ \emph {et~al.}(2023)\citenamefont {Zawala}, \citenamefont {Miguet}, \citenamefont {Rastogi}, \citenamefont {Atasi}, \citenamefont {Borkowski}, \citenamefont {Scheid},\ and\ \citenamefont {Fuller}}]{ZMRABSF23}%
  \BibitemOpen
  \bibfield  {author} {\bibinfo {author} {\bibfnamefont {J.}~\bibnamefont {Zawala}}, \bibinfo {author} {\bibfnamefont {J.}~\bibnamefont {Miguet}}, \bibinfo {author} {\bibfnamefont {P.}~\bibnamefont {Rastogi}}, \bibinfo {author} {\bibfnamefont {O.}~\bibnamefont {Atasi}}, \bibinfo {author} {\bibfnamefont {M.}~\bibnamefont {Borkowski}}, \bibinfo {author} {\bibfnamefont {B.}~\bibnamefont {Scheid}}, \ and\ \bibinfo {author} {\bibfnamefont {G.~G.}\ \bibnamefont {Fuller}},\ }\bibfield  {title} {\enquote {\bibinfo {title} {Coalescence of surface bubbles: The crucial role of motion-induced dynamic adsorption layer},}\ }\href@noop {} {\bibfield  {journal} {\bibinfo  {journal} {Adv. Colloid Interface Sci.}\ }\textbf {\bibinfo {volume} {317}},\ \bibinfo {pages} {102916} (\bibinfo {year} {2023})}\BibitemShut {NoStop}%
\bibitem [{\citenamefont {Li}\ and\ \citenamefont {Mao}(2001)}]{LM01}%
  \BibitemOpen
  \bibfield  {author} {\bibinfo {author} {\bibfnamefont {X.-J.}\ \bibnamefont {Li}}\ and\ \bibinfo {author} {\bibfnamefont {Z.-S.}\ \bibnamefont {Mao}},\ }\bibfield  {title} {\enquote {\bibinfo {title} {The effect of surfactant on the motion of a buoyancy-driven drop at intermediate reynolds numbers: A numerical approach},}\ }\href@noop {} {\bibfield  {journal} {\bibinfo  {journal} {J. Colloid Interface Sci.}\ }\textbf {\bibinfo {volume} {240}},\ \bibinfo {pages} {307--322} (\bibinfo {year} {2001})}\BibitemShut {NoStop}%
\bibitem [{\citenamefont {Palaparthi}\ \emph {et~al.}(2006)\citenamefont {Palaparthi}, \citenamefont {Papageorgiou},\ and\ \citenamefont {Maldarelli}}]{PPM06}%
  \BibitemOpen
  \bibfield  {author} {\bibinfo {author} {\bibfnamefont {R.}~\bibnamefont {Palaparthi}}, \bibinfo {author} {\bibfnamefont {D.}~\bibnamefont {Papageorgiou}}, \ and\ \bibinfo {author} {\bibfnamefont {C.}~\bibnamefont {Maldarelli}},\ }\bibfield  {title} {\enquote {\bibinfo {title} {Theory and experiments on the stagnant cap regime in the motion of spherical surfactant-laden bubbles},}\ }\href@noop {} {\bibfield  {journal} {\bibinfo  {journal} {J. Fluid Mech.}\ }\textbf {\bibinfo {volume} {559}},\ \bibinfo {pages} {1--44} (\bibinfo {year} {2006})}\BibitemShut {NoStop}%
\bibitem [{\citenamefont {Takagi}\ \emph {et~al.}(2003)\citenamefont {Takagi}, \citenamefont {Uda}, \citenamefont {Watanabe},\ and\ \citenamefont {Matsumoto}}]{TUWM03}%
  \BibitemOpen
  \bibfield  {author} {\bibinfo {author} {\bibfnamefont {S.}~\bibnamefont {Takagi}}, \bibinfo {author} {\bibfnamefont {T.}~\bibnamefont {Uda}}, \bibinfo {author} {\bibfnamefont {Y.}~\bibnamefont {Watanabe}}, \ and\ \bibinfo {author} {\bibfnamefont {Y.}~\bibnamefont {Matsumoto}},\ }\bibfield  {title} {\enquote {\bibinfo {title} {Behavior of a rising bubble in water with surfactant dissolution (1st report, steady behavior) (in japanese) .}}\ }\href@noop {} {\bibfield  {journal} {\bibinfo  {journal} {Trans. JSME B}\ }\textbf {\bibinfo {volume} {69}},\ \bibinfo {pages} {2192–2199} (\bibinfo {year} {2003})}\BibitemShut {NoStop}%
\bibitem [{\citenamefont {Pesci}\ \emph {et~al.}(2018)\citenamefont {Pesci}, \citenamefont {Weiner}, \citenamefont {Marschall},\ and\ \citenamefont {Bothe}}]{PWMB18}%
  \BibitemOpen
  \bibfield  {author} {\bibinfo {author} {\bibfnamefont {C.}~\bibnamefont {Pesci}}, \bibinfo {author} {\bibfnamefont {A.}~\bibnamefont {Weiner}}, \bibinfo {author} {\bibfnamefont {H.}~\bibnamefont {Marschall}}, \ and\ \bibinfo {author} {\bibfnamefont {D.}~\bibnamefont {Bothe}},\ }\bibfield  {title} {\enquote {\bibinfo {title} {Computational analysis of single rising bubbles influenced by soluble surfactant},}\ }\href@noop {} {\bibfield  {journal} {\bibinfo  {journal} {J. Fluid Mech.}\ }\textbf {\bibinfo {volume} {856}},\ \bibinfo {pages} {709--763} (\bibinfo {year} {2018})}\BibitemShut {NoStop}%
\bibitem [{\citenamefont {Tagawa}\ \emph {et~al.}(2014)\citenamefont {Tagawa}, \citenamefont {Takagi},\ and\ \citenamefont {Matsumoto}}]{TTM14}%
  \BibitemOpen
  \bibfield  {author} {\bibinfo {author} {\bibfnamefont {Y.}~\bibnamefont {Tagawa}}, \bibinfo {author} {\bibfnamefont {S.}~\bibnamefont {Takagi}}, \ and\ \bibinfo {author} {\bibfnamefont {Y.}~\bibnamefont {Matsumoto}},\ }\bibfield  {title} {\enquote {\bibinfo {title} {Surfactant effect on path instability of a rising bubble},}\ }\href@noop {} {\bibfield  {journal} {\bibinfo  {journal} {J. Fluid Mech.}\ }\textbf {\bibinfo {volume} {738}},\ \bibinfo {pages} {124--142} (\bibinfo {year} {2014})}\BibitemShut {NoStop}%
\bibitem [{\citenamefont {Manikantan}\ and\ \citenamefont {Squires}(2020)}]{MS20}%
  \BibitemOpen
  \bibfield  {author} {\bibinfo {author} {\bibfnamefont {H.}~\bibnamefont {Manikantan}}\ and\ \bibinfo {author} {\bibfnamefont {T.~M.}\ \bibnamefont {Squires}},\ }\bibfield  {title} {\enquote {\bibinfo {title} {Surfactant dynamics: hidden variables controlling fluid flows},}\ }\href@noop {} {\bibfield  {journal} {\bibinfo  {journal} {J. Fluid Mech.}\ }\textbf {\bibinfo {volume} {892}},\ \bibinfo {pages} {P1} (\bibinfo {year} {2020})}\BibitemShut {NoStop}%
\bibitem [{\citenamefont {Varghese}\ \emph {et~al.}(2024)\citenamefont {Varghese}, \citenamefont {Sykes}, \citenamefont {Quetzeri-Santiago}, \citenamefont {Castrej\'on-Pita},\ and\ \citenamefont {Castrej\'on-Pita}}]{VSQCC24}%
  \BibitemOpen
  \bibfield  {author} {\bibinfo {author} {\bibfnamefont {N.}~\bibnamefont {Varghese}}, \bibinfo {author} {\bibfnamefont {T.~C.}\ \bibnamefont {Sykes}}, \bibinfo {author} {\bibfnamefont {M.~A.}\ \bibnamefont {Quetzeri-Santiago}}, \bibinfo {author} {\bibfnamefont {A.~A.}\ \bibnamefont {Castrej\'on-Pita}}, \ and\ \bibinfo {author} {\bibfnamefont {J.~R.}\ \bibnamefont {Castrej\'on-Pita}},\ }\bibfield  {title} {\enquote {\bibinfo {title} {Effect of surfactants on the splashing dynamics of drops impacting smooth substrates},}\ }\href@noop {} {\bibfield  {journal} {\bibinfo  {journal} {Langmuir}\ }\textbf {\bibinfo {volume} {40}},\ \bibinfo {pages} {8781--8790} (\bibinfo {year} {2024})}\BibitemShut {NoStop}%
\bibitem [{\citenamefont {Ponce-Torres}\ \emph {et~al.}(2020)\citenamefont {Ponce-Torres}, \citenamefont {Rubio}, \citenamefont {Herrada}, \citenamefont {Eggers},\ and\ \citenamefont {Montanero}}]{PRHEM20}%
  \BibitemOpen
  \bibfield  {author} {\bibinfo {author} {\bibfnamefont {A.}~\bibnamefont {Ponce-Torres}}, \bibinfo {author} {\bibfnamefont {M.}~\bibnamefont {Rubio}}, \bibinfo {author} {\bibfnamefont {M.~A.}\ \bibnamefont {Herrada}}, \bibinfo {author} {\bibfnamefont {J.}~\bibnamefont {Eggers}}, \ and\ \bibinfo {author} {\bibfnamefont {J.~M.}\ \bibnamefont {Montanero}},\ }\bibfield  {title} {\enquote {\bibinfo {title} {Influence of the surface viscous stress on the pinch-off of free surfaces loaded with nearly-inviscid surfactants},}\ }\href@noop {} {\bibfield  {journal} {\bibinfo  {journal} {Sci. Rep.}\ }\textbf {\bibinfo {volume} {10}},\ \bibinfo {pages} {16065} (\bibinfo {year} {2020})}\BibitemShut {NoStop}%
\bibitem [{\citenamefont {Craster}\ \emph {et~al.}(2009)\citenamefont {Craster}, \citenamefont {Matar},\ and\ \citenamefont {Papageorgiou}}]{CMP09}%
  \BibitemOpen
  \bibfield  {author} {\bibinfo {author} {\bibfnamefont {R.~V.}\ \bibnamefont {Craster}}, \bibinfo {author} {\bibfnamefont {O.~K.}\ \bibnamefont {Matar}}, \ and\ \bibinfo {author} {\bibfnamefont {D.~T.}\ \bibnamefont {Papageorgiou}},\ }\bibfield  {title} {\enquote {\bibinfo {title} {Breakup of surfactant-laden jets above the critical micelle concentration},}\ }\href@noop {} {\bibfield  {journal} {\bibinfo  {journal} {J. Fluid Mech.}\ }\textbf {\bibinfo {volume} {629}},\ \bibinfo {pages} {195--219} (\bibinfo {year} {2009})}\BibitemShut {NoStop}%
\bibitem [{\citenamefont {Kalogirou}\ and\ \citenamefont {Blyth}(2019)}]{KB19}%
  \BibitemOpen
  \bibfield  {author} {\bibinfo {author} {\bibfnamefont {A.}~\bibnamefont {Kalogirou}}\ and\ \bibinfo {author} {\bibfnamefont {M.~G.}\ \bibnamefont {Blyth}},\ }\bibfield  {title} {\enquote {\bibinfo {title} {The role of soluble surfactants in the linear stability of two-layer flow in a channel},}\ }\href@noop {} {\bibfield  {journal} {\bibinfo  {journal} {J. Fluid Mech.}\ }\textbf {\bibinfo {volume} {873}},\ \bibinfo {pages} {18--48} (\bibinfo {year} {2019})}\BibitemShut {NoStop}%
\bibitem [{\citenamefont {Tricot}(1997)}]{T97}%
  \BibitemOpen
  \bibfield  {author} {\bibinfo {author} {\bibfnamefont {Y.-M.}\ \bibnamefont {Tricot}},\ }\enquote {\bibinfo {title} {{Surfactants: Static and Dynamic Surface Tension}},}\ \ (\bibinfo  {publisher} {Chapman and Hall},\ \bibinfo {year} {1997})\ pp.\ \bibinfo {pages} {100--136}\BibitemShut {NoStop}%
\bibitem [{\citenamefont {Theofilis}(2011)}]{T11}%
  \BibitemOpen
  \bibfield  {author} {\bibinfo {author} {\bibfnamefont {V.}~\bibnamefont {Theofilis}},\ }\bibfield  {title} {\enquote {\bibinfo {title} {Global linear instability},}\ }\href@noop {} {\bibfield  {journal} {\bibinfo  {journal} {Annu. Rev. Fluid Mech.}\ }\textbf {\bibinfo {volume} {43}},\ \bibinfo {pages} {319--352} (\bibinfo {year} {2011})}\BibitemShut {NoStop}%
\bibitem [{\citenamefont {Herrada}\ and\ \citenamefont {Montanero}(2016)}]{HM16a}%
  \BibitemOpen
  \bibfield  {author} {\bibinfo {author} {\bibfnamefont {M.~A}\ \bibnamefont {Herrada}}\ and\ \bibinfo {author} {\bibfnamefont {J.~M.}\ \bibnamefont {Montanero}},\ }\bibfield  {title} {\enquote {\bibinfo {title} {A numerical method to study the dynamics of capillary fluid systems},}\ }\href@noop {} {\bibfield  {journal} {\bibinfo  {journal} {J. Comput. Phys.}\ }\textbf {\bibinfo {volume} {306}},\ \bibinfo {pages} {137--147} (\bibinfo {year} {2016})}\BibitemShut {NoStop}%
\bibitem [{\citenamefont {Herrada}(2023)}]{JAM}%
  \BibitemOpen
  \bibfield  {author} {\bibinfo {author} {\bibfnamefont {M.~A.}\ \bibnamefont {Herrada}},\ }\bibfield  {title} {\enquote {\bibinfo {title} {{This method has recently been termed JAM (Jacobian Analytical Method). Examples of JAM codes can be found at https://github.com/miguelherrada/JAM.}}}\ }\href@noop {} {\  (\bibinfo {year} {2023})}\BibitemShut {NoStop}%
\bibitem [{\citenamefont {Canny}(1986)}]{C86}%
  \BibitemOpen
  \bibfield  {author} {\bibinfo {author} {\bibfnamefont {J.}~\bibnamefont {Canny}},\ }\bibfield  {title} {\enquote {\bibinfo {title} {A computational approach to edge-detection},}\ }\href@noop {} {\bibfield  {journal} {\bibinfo  {journal} {IEEE Trans. Pattern Anal. Mach. Intell.}\ }\textbf {\bibinfo {volume} {8}},\ \bibinfo {pages} {679--698} (\bibinfo {year} {1986})}\BibitemShut {NoStop}%
\bibitem [{\citenamefont {Penkina}\ \emph {et~al.}(2016)\citenamefont {Penkina}, \citenamefont {Zolnikov}, \citenamefont {Pomazyonkova},\ and\ \citenamefont {Avramenko}}]{PZPA16}%
  \BibitemOpen
  \bibfield  {author} {\bibinfo {author} {\bibfnamefont {Y.~A.}\ \bibnamefont {Penkina}}, \bibinfo {author} {\bibfnamefont {I.~M.}\ \bibnamefont {Zolnikov}}, \bibinfo {author} {\bibfnamefont {A.~E.}\ \bibnamefont {Pomazyonkova}}, \ and\ \bibinfo {author} {\bibfnamefont {G.~V.}\ \bibnamefont {Avramenko}},\ }\bibfield  {title} {\enquote {\bibinfo {title} {Colloid-chemical properties of nonionic gemini surfactants {Surfynol} 400 series with different degrees of oxyethylation},}\ }\href@noop {} {\bibfield  {journal} {\bibinfo  {journal} {Butlerov Communications}\ }\textbf {\bibinfo {volume} {46}},\ \bibinfo {pages} {92--101} (\bibinfo {year} {2016})}\BibitemShut {NoStop}%
\bibitem [{\citenamefont {Tajima}\ \emph {et~al.}(1970)\citenamefont {Tajima}, \citenamefont {Muramatsu},\ and\ \citenamefont {Sasaki}}]{TMS70}%
  \BibitemOpen
  \bibfield  {author} {\bibinfo {author} {\bibfnamefont {K.}~\bibnamefont {Tajima}}, \bibinfo {author} {\bibfnamefont {M.}~\bibnamefont {Muramatsu}}, \ and\ \bibinfo {author} {\bibfnamefont {T.}~\bibnamefont {Sasaki}},\ }\bibfield  {title} {\enquote {\bibinfo {title} {Radiotracer studies on adsorption of surface active substance at aqueous surface. {I}. {Accurate} measurement of adsorption of {Tritiated Sodium Dodecylsulfate}},}\ }\href@noop {} {\bibfield  {journal} {\bibinfo  {journal} {Bull. Chem. Soc. Japan}\ }\textbf {\bibinfo {volume} {43}},\ \bibinfo {pages} {1991--1998} (\bibinfo {year} {1970})}\BibitemShut {NoStop}%
\bibitem [{\citenamefont {Jenny}\ \emph {et~al.}(2003)\citenamefont {Jenny}, \citenamefont {Bouchet},\ and\ \citenamefont {Dusek}}]{JBD03}%
  \BibitemOpen
  \bibfield  {author} {\bibinfo {author} {\bibfnamefont {M.}~\bibnamefont {Jenny}}, \bibinfo {author} {\bibfnamefont {G.}~\bibnamefont {Bouchet}}, \ and\ \bibinfo {author} {\bibfnamefont {J.}~\bibnamefont {Dusek}},\ }\bibfield  {title} {\enquote {\bibinfo {title} {Nonvertical ascension or fall of a free sphere in a newtonian fluid},}\ }\href@noop {} {\bibfield  {journal} {\bibinfo  {journal} {Phys. Fluids}\ }\textbf {\bibinfo {volume} {15}},\ \bibinfo {pages} {L9--L12} (\bibinfo {year} {2003})}\BibitemShut {NoStop}%
\bibitem [{\citenamefont {Jenny}\ \emph {et~al.}(2004)\citenamefont {Jenny}, \citenamefont {Dusek},\ and\ \citenamefont {Bouchet}}]{JDB04}%
  \BibitemOpen
  \bibfield  {author} {\bibinfo {author} {\bibfnamefont {M.}~\bibnamefont {Jenny}}, \bibinfo {author} {\bibfnamefont {J.}~\bibnamefont {Dusek}}, \ and\ \bibinfo {author} {\bibfnamefont {G.}~\bibnamefont {Bouchet}},\ }\bibfield  {title} {\enquote {\bibinfo {title} {Instabilities and transition of a sphere falling or ascending in a {Newtonian} fluid},}\ }\href@noop {} {\bibfield  {journal} {\bibinfo  {journal} {J. Fluid Mech.}\ }\textbf {\bibinfo {volume} {508}},\ \bibinfo {pages} {201--239} (\bibinfo {year} {2004})}\BibitemShut {NoStop}%
\bibitem [{\citenamefont {Clift}\ \emph {et~al.}(1978)\citenamefont {Clift}, \citenamefont {Grace},\ and\ \citenamefont {Weber}}]{CGW78}%
  \BibitemOpen
  \bibfield  {author} {\bibinfo {author} {\bibfnamefont {R.}~\bibnamefont {Clift}}, \bibinfo {author} {\bibfnamefont {J.~R.}\ \bibnamefont {Grace}}, \ and\ \bibinfo {author} {\bibfnamefont {M.~E.}\ \bibnamefont {Weber}},\ }\href@noop {} {\emph {\bibinfo {title} {Bubbles, Drops and Particles}}}\ (\bibinfo  {publisher} {Academic Press},\ \bibinfo {address} {USA},\ \bibinfo {year} {1978})\BibitemShut {NoStop}%
\bibitem [{\citenamefont {Mei}\ \emph {et~al.}(1994)\citenamefont {Mei}, \citenamefont {Klausner},\ and\ \citenamefont {Lawrence}}]{MKL94}%
  \BibitemOpen
  \bibfield  {author} {\bibinfo {author} {\bibfnamefont {R.}~\bibnamefont {Mei}}, \bibinfo {author} {\bibfnamefont {J.~F.}\ \bibnamefont {Klausner}}, \ and\ \bibinfo {author} {\bibfnamefont {C.~J.}\ \bibnamefont {Lawrence}},\ }\bibfield  {title} {\enquote {\bibinfo {title} {A note on the history force on a spherical bubble at finite reynolds number},}\ }\href@noop {} {\bibfield  {journal} {\bibinfo  {journal} {Phys. Fluids}\ }\textbf {\bibinfo {volume} {6}},\ \bibinfo {pages} {418--420} (\bibinfo {year} {1994})}\BibitemShut {NoStop}%
\end{thebibliography}

%

\end{document}